\documentclass{ws-ijmpe}
\usepackage[super,compress]{cite}
\usepackage{graphicx}
\usepackage{amsmath}
\usepackage{amssymb}
\usepackage{mathtools}
\usepackage{liealg}
\usepackage{wignert}
\usepackage{isotope}
\usepackage{dcolumn}
\usepackage{mathrsfs}

\DeclareRobustCommand{\Lambdahat}{\hat{\Lambda}}

\DeclareRobustCommand{\scrD}{\mathscr{D}}

\begin{document}

\markboth{S. Frauendorf}{The Low-Energy Quadrupole Mode of Nuclei}

\catchline{}{}{}{}{}
%%%%%%%%%%%%%%%%%%%%%%%%%%%%%%%%%%%%%%%%%%%%%%%%%%%%%%%%%%%%%%%%%%%%

\title{THE LOW-ENERGY QUADRUPOLE MODE OF NUCLEI}

\author{\footnotesize S. FRAUENDORF}

\address{Department of Physics, University Notre Dame,\\
Notre Dame, IN 46556, USA\\
sfrauend@nd.edu}
\maketitle

\begin{history}
\received{Day Month Year}
\revised{Day Month Year}
%\accepted{Day Month Year}
%\comby{(xxxxxxxxxx)}
\end{history}

\begin{abstract}
The phenomenological classification of collective quadrupole excitations by means of the Bohr Hamiltonian is reviewed with focus on signatures for
triaxility. The variants of the microscopic Bohr Hamiltonian derived by means of the Adiabatic Time Dependent Mean Field theory from the Pairing plus Quadrupole-Quadrupole
interaction, the Shell Correction Method, the Skyrme Energy Density Functional, the Relativistic Mean Field Theory, and the Gogny interaction are discussed and 
applications to concrete nuclides reviewed. The Generator Coordinate Method for the five dimensional quadrupole deformation space and first applications 
to triaxial nuclei are presented.   The phenomenological classification in the framework of the Interacting Boson Model is discussed with a critical view on the boson number 
counting rule.  The recent success in calculating the model parameters by mapping the mean field deformation energy surface on the bosonic  one is discussed and the 
applications listed.   A critical assessment of the models is given with focus on the limitations due to the adiabatic approximation.  
The Tidal Wave approach and the Triaxial Projected Shell Model are presented as practical approaches to calculate spectral properties outside the adiabatic region.

\end{abstract}


\ccode{21.10.Re, 21.60.Ev, 21.60.Fw, 23.20.Lv, 24.10.Cn}

%\tableofcontents

\section{Introduction}
Nuclei with mass number $A>50$ are compact systems with a well defined surface, the shape of which and its orientation 
in space constitute collective degrees of freedom, which are discussed in any  textbook on nuclear structure. This article 
reviews the descriptions of the quadruple mode and tries to asses,
to which extend present nuclear theory is capable of describing and predicting the properties of the corresponding quantum states on the basis
of the underpinning degrees of freedom of the nucleonic constituents. 
       There are two major phenomenological descriptions of the collective quadrupole modes at low spin: the 
Bohr - Hamiltonian (BH),  and the Interacting Boson Model (IBM), which describes the quadrupole mode in terms of 
boson operators that form a closed  SU(6) Lie algebra.
% The two models have been generalized to include the collective octupole modes as well.  
Both approaches are well exposed in textbooks, as for example \cite{BMII,EG,RW,IA87,Casten}. 
Section \ref{sec:GCMphen} reviews some basics of the description in terms of the BH 
and recent developments of this phenomenology, where the aspects triaxiality are exposed in some detail.
A microscopic version of the BH has been derived in the framework of all mean field approaches used in practice at present.
Section \ref{sec:ATDMF}  gives an overview of derivations in the framework of the various  
Adiabatic Time-Dependent Mean Field approaches. Section \ref{sec:GCM} covers corresponding work in the framework of the
Generator Coordinate Method. Applications to concrete nuclei are listed and selected examples discussed. Section \ref{sec:IBM}
presents some basics of the IBM phenomenology  and the recent successful calculation of its  parameters by means of 
mapping the  potential energy surface of the mean field.      

\begin{figure}[t]
\psfig{file=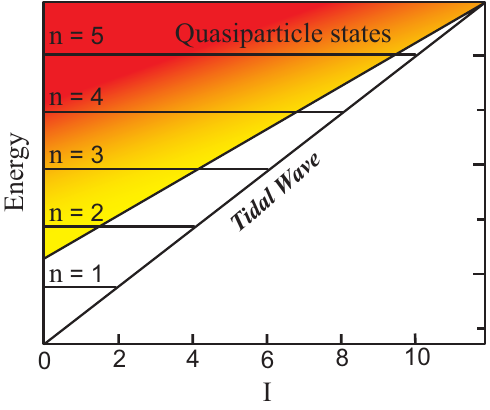,width=5.9cm}
\psfig{file=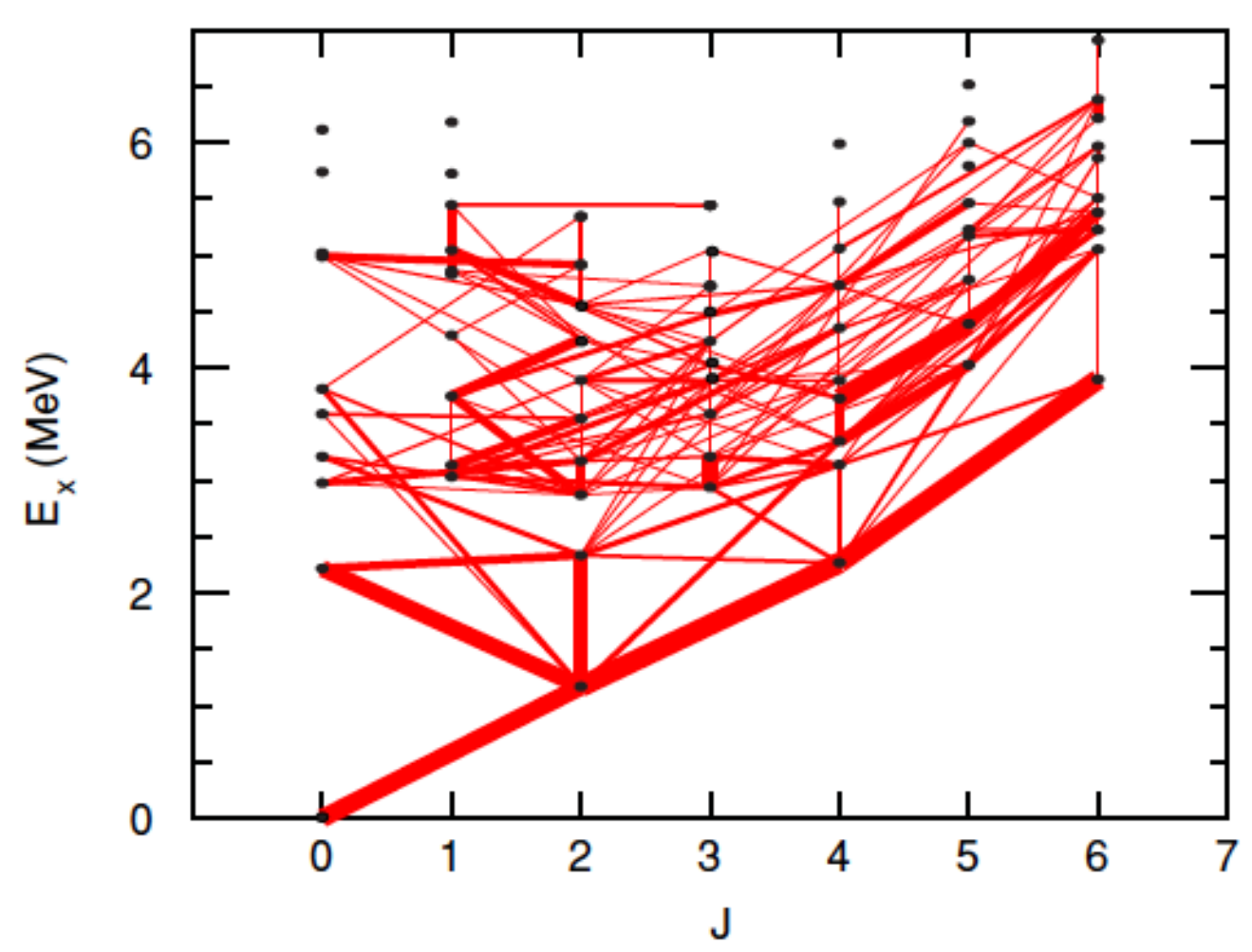,width=6.6cm}
%\vspace*{-1cm}
\caption{ Left: Schematic representation of the location of the collective quadrupole vibrational excitations relative to the quasiparticle excitations. 
The darker shades approximately indicate higher densities of quasiparticle states.  From Ref. \cite{102PdTidalPRL}.
Right: Shell Model Calculation for $^{62}$Ni. The width of the bars is proportional to the $B(E2)$ value of the connecting transition. From Ref. \cite{Chakraborty11}.}
\label{fig:decoherence}
\end{figure}

Nuclei are composed of a relative small number number of nucleons 
compared to other many-body systems. As a consequence, the "granular structure" of the collective degrees of freedom appears already  after 
the excitation of few quanta, which results in a progressive decoherence of the collective modes. The left side of
Fig. \ref{fig:decoherence} illustrates the point in a schematic way for a vibrational nucleus: 
The multi-phonon excitations encounter very soon the region of the quasiparticle excitations to which they couple. The right side shows a realistic Shell Model
calculation. In accordance with the white adiabatic region on the left side, the collectively enhanced transitions are restricted  to the one and two phonon 
states and higher up to the yrast region.    
The purely collective models covered in sections \ref{sec:BH} - \ref{sec:IBM} assume adiabaticity of the collective motion explicitly or implicitly.
Their realm is restricted to the  white adiabatic area in Fig. \ref{fig:decoherence}. Outside, the coupling between the quasiparticle and collective degrees
of freedom must be taken into account in a non-perturbative way. 
Section \ref{sec:noadia} discuss the Tidal Wave approach and the Triaxial Projected Shell Model, which
take this coupling into account. Section \ref{sec:assessment} tries to assess the presently used models and points out some challenges, which 
seem important from the author's point of view. 

%The small size of nuclei allows us to classify the discrete nuclear states by their angular momentum. The nucleus may acquire  angular momentum 
%in a coherent way, which results in regularly spaced rotational  bands, or in an incoherent way, by aligning the angular momenta of individual 
%quasiparticles. I will discuss examples of the resulting  interplay between these "single-particle and rotational" degrees of freedom. The standard
 %point of view of textbooks, is to consider
%collective rotation  as the change of the orientation of the deformed nuclear shape. 
%Nuclear orientation and the ensuing appearance of rotational bands may arise from more general types unisotropy of the nucleonic motion.
%Their nature and the new forms of rotational motion that they generate  will be discussed. 

%Other collective modes, like the  Giant Resonances, the Scissors Mode, and the Dipole Pygmy Resonance
%will not be considered, because they have the character of driven oscillations.  This article focuses on self-sustianed collective modes.                  

\section{The Bohr Hamiltonian}\label{sec:BH}

The Bohr coordinate system and Hamiltonian are reviewed in detail by Prochniak and Rohozinski~\cite{prochniak2009}.  
The collective states are represented by wave functions of the components 
$\alpha_\mu$ ($\mu=-2$, $\ldots$, $2$) of the scale-free quadrupole deformation tensor 
which are expressed by the five-dimensional spherical polar coordinates 
\begin{equation}\label{eqn-bohr-q}
\alpha_\mu = \beta \left[ \cos\gamma\, \scrD^{(2)}_{0,M}(\Omega) +
\frac{1}{\sqrt{2}} \sin\gamma \left[ \scrD^{(2)}_{2,M}(\Omega) +
\scrD^{(2)}_{-2,M}(\Omega) \right] \right],
\end{equation}
where $\Omega$ are the Euler angles specifying the orientation of the shape.

\subsection{The Geometric Collective Model}\label{sec:GCMphen}

The Geometric Collective Model (GCM) is a parametrized version of the BH based on an expansion into scalars 
of increasing power constructed from the coordinate $\alpha_\mu$ and their conjugate momenta $\pi_\mu$, which was
introduced by Gneuss and Greiner \cite{GG71}. Only the quadratic term in $\pi_\mu$ is kept.
 In terms of the
quadrupole deformation variables $\beta$ and $\gamma$ and Euler angles $\Omega$, 
the Bohr Hamiltonian is given by
\begin{equation}
\label{eqn-HGCM2}
H_{GCM2}=\frac{\hbar^2}{\sqrt{5}B_2}\biggl[
T_{\beta\beta}+
\frac{\Lambdahat^2}{\beta^2}\biggr]+V(\beta,\gamma),
\end{equation}
where
\begin{equation}\label{eqn-Tbb}
T_{\beta\beta}=-\frac{1}{\beta^4}\frac{\partial}{\partial\beta}\beta^4\frac{\partial}{\partial\beta}
\end{equation} 
and
\begin{equation}
\label{eqn-Lambdasqr}
\Lambdahat^2=-\biggl(
\frac{1}{\sin 3\gamma} 
\frac{\partial}{\partial \gamma} \sin 3\gamma \frac{\partial}{\partial \gamma}
- \frac{1}{4}
\sum_{i=1,2,3} \frac{\hat{L}_i^{\prime2}}{\sin^2(\gamma -\frac{2}{3} \pi i)}
\biggr).
\end{equation}
The operator appearing in brackets in the kinetic energy is the
Laplacian in five dimensions.  Its angular part $\Lambdahat^2$ is
the Casimir operator for the five-dimensional rotation group
$\grpso{5}$, 
which contains the rotations in physical space, acting on
the Euler angle coordinates, as an $\grpso{3}$ subgroup. 
The potential energy
$V(\beta,\gamma)$ must be periodic in $\gamma$, with period
$120^\circ$, and it must be symmetric about $\gamma=0^\circ$ and
$\gamma=60^\circ$.  
 Prochniak and Rohozinski \cite{prochniak2009} discuss in detail the various methods for calculating the eigenvalues and eigenfunctions of the BH.

\begin{figure}[th]
\centerline{\psfig{file=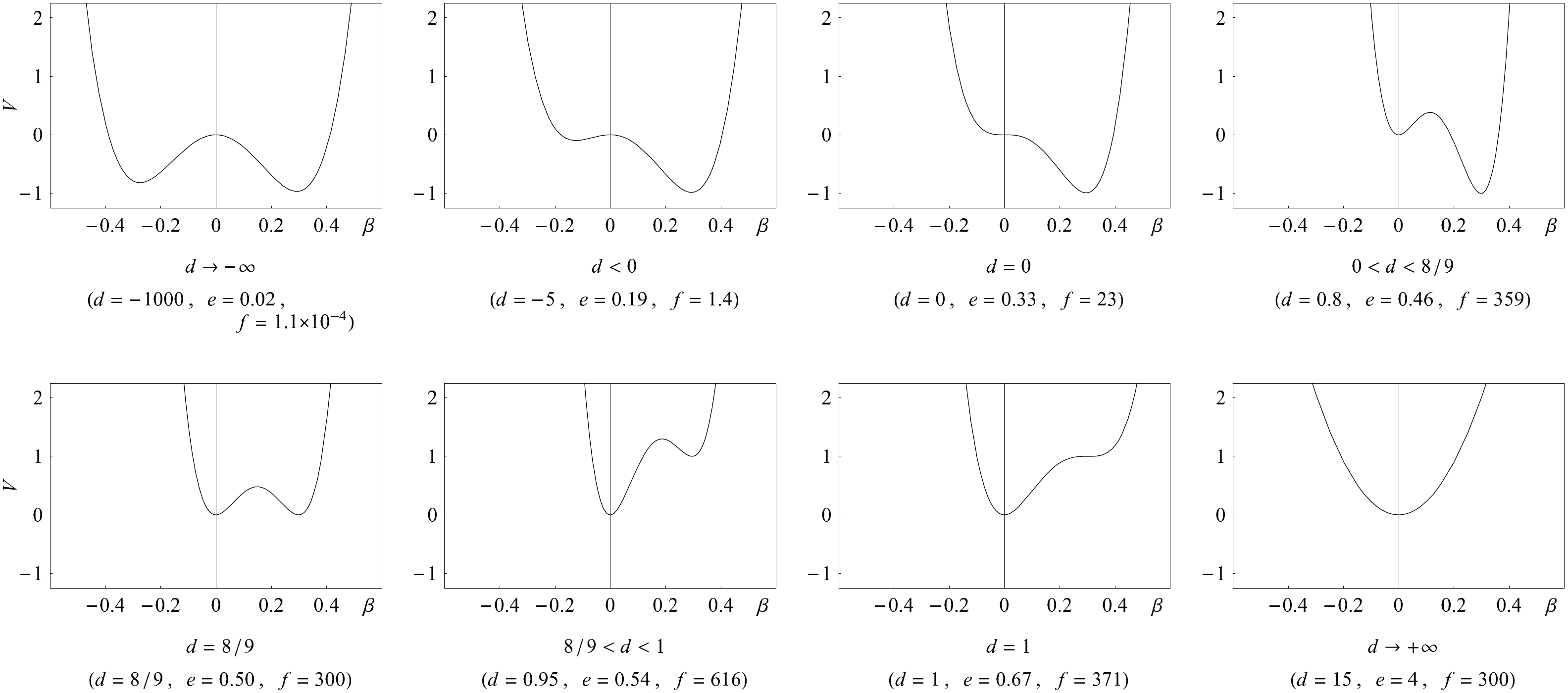,width=\linewidth}}
\caption{Illustration of the qualitatively different shapes of the GCM potential function (\ref{eqn-VGCM2}) obtained for different ranges of values for the
parameter $d$. Potentials are shown as a function of $\beta$, where $\gamma=0$ ($\beta<0$ is equivalent with $\beta>0$, $\gamma=\pi/3$). 
For $f$ in MeV, the energy scale is also in MeV. Reproduced from Ref. \cite{Caprio2par}.
\label{fig:gcm_pot}}
\end{figure}

The properties of the experimental  $2^+_1$, $4^+_1$, $6^+_1$, $2^+_2$, $3^+_1$, $4^+_2$, $0^+_2$, $4^+_3$ states can be classified
by assuming that the collective potential contains only three terms,
\begin{equation}\label{eqn-VGCM2}
V_{A}(\beta,\gamma)=\frac{1}{\sqrt{5}}C_2\beta^2-\sqrt{\frac{2}{35}}C_3\beta^3\cos3\gamma+\frac{1}{5}C_4\beta^4.
\end{equation}   
This provides a more complete scheme than the traditional classification into rotational and vibrational nuclei.  
Caprio \cite{Caprio2par} demonstrated that the structure of the collective wave function is determined by  two dimensionless parameter ratios only, 
 the control parameter $d=112C_2C_4/(9\sqrt{5}C_3)$ of the potential and the structure parameter $\hbar^2S=\hbar^2C_4^5/(C_3^6B_2)$, which controls the zero-point energy.  
The scale of the deformation parameter $\beta$ is fixed by the ratio $e=C_3/C_4$. The scale of the total energy is $g=\hbar^2/(\sqrt{5}e^2B_2)$. 
For illustration, it is useful to introduce the scale of the potential energy $f=C_4e^4=C_3^4/C_4^3$. Then the structure parameter $S=f/g$
is the ratio between the scales of the potential and the total energy. It controls to which extend the zero-point fluctuations of $\beta$ and $\gamma$
wash out the details of the potential.    In scaled form, the GCM Hamiltonian reads
\begin{equation}\label{eqn-HGCM2s}
H_{GCM2}=g\biggl[
T_{\bar\beta\bar\beta}+
\frac{\Lambdahat^2}{\bar\beta^2}+\frac{\sqrt{5}}{\hbar^2S}\left(\frac{9}{112}d\bar\beta^2-\sqrt{\frac{2}{35}}\bar\beta^3\cos3\gamma+\frac{1}{5}\bar\beta^4\right)\biggr],
\end{equation}
where $\bar\beta=\beta/e$ is the scale-free deformation parameter. 
The matrix elements of the charge quadrupole moments, which generate the E2 $\gamma$-transitions and the static electric quadrupole 
matrix elements, are modeled by a homogeneously charged droplet, 
\begin{equation}
Q_\mu=\frac{3ZR_0^2}{4\pi}\left(e\bar\alpha_\mu^*-\frac{10}{\sqrt{70\pi}}e^2[\bar\alpha\times\bar\alpha]^{(2)*}_\mu\right),
\end{equation}
which fixes the scale $e$.

\begin{figure}[th]
\centerline{\psfig{file=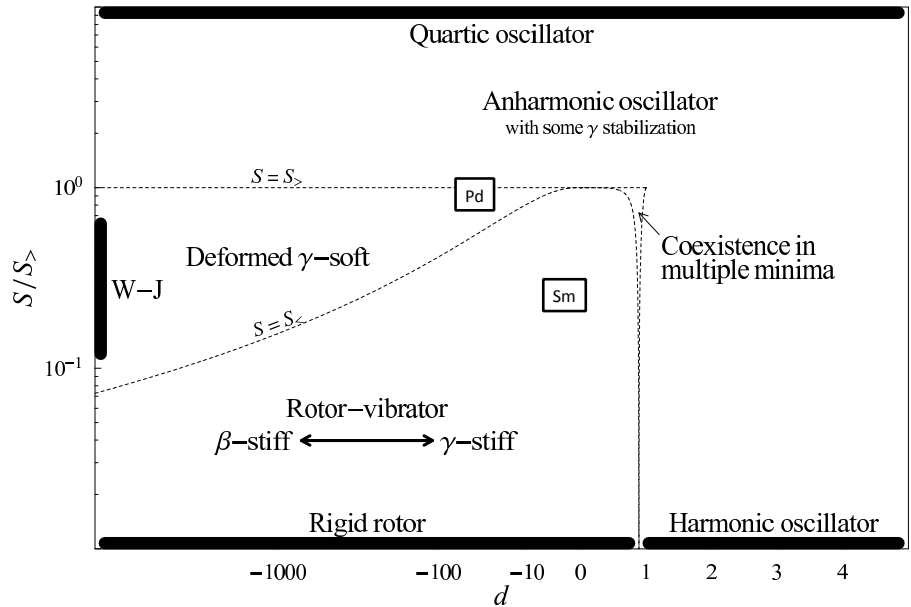,width=14cm}}
\vspace*{-1cm}
\caption{Map of the GCM ($d$, $S$) parameter
space. The regions in which qualitatively different
structures occur are indicated. The curves $S=S_<$, and $S=S_>$ (dotted lines) provide estimates
for the approximate boundaries between these regions. 
The squares show the location of $^{102}$Pd and $^{152}$Sm in the parameter space. 
Bars along the edges of the plot represent
structures which occur in their ideal form
at $d\rightarrow±\infty$ or at $S=0$ or $\infty$. "W-J" denotes the
Wilets-Jean rigidly-deformed $\gamma$-soft structure. Reproduced  and modified from Ref. \cite{Caprio2par}.
\label{fig:GCMmap}}
\end{figure}

Fig. \ref{fig:gcm_pot} illustrates the different types of potentials.   For $0<d<1$ the potential has two minima at prolate deformation.
For $d<0$ there is a saddle at oblate shape, which
 connects smoothly with the prolate minimum via the $\gamma$ degree of freedom.
For $d\rightarrow -\infty$ the limit of  $\gamma$-independence is approached. 
 Caprio carried out a qualitative analysis of the ground state  wave function
 by means of the WKBA. Comparing the zero-point energy with the extrema of the potential leads to a classification of 
 structure of the low-lying collective states, which is mapped in  Fig. \ref{fig:GCMmap}. 
 
At the line   $S_>$, the zero-point energy is equal to the height of the barrier between the two axial minima. 
Above this line the wave function spreads over both minima.
Below the line   $S_>$, the wave function becomes progressively suppressed under the barrier. 
That is, the region of deformed nuclei lies within the rectangle $[d<1$, $S_>]$. 
 At the line $S_<$, the zero-point energy is equal to the 
energy difference between the oblate saddle and prolate minimum. That is, the
prolate nuclei are located  below the line $S_<$, which is denoted by  "rotor-vibrator" region.    The "stiffness"
for the $\beta$- and $\gamma$-vibrations varies with $d$ (double arrow), where "stiffness" refers to the relative order
of the $0^+_2$ and $2^+_2$ states. The lines $S_>$ and $S_<$ demarcate the region of $\gamma$-soft
nuclei, for which the wave function extends over the whole range $0\leq\gamma\leq\pi/3$.
Outside the rectangle  $[d<1$, $S_>]$, the wave function changes gradually from a harmonic oscillator to a quartic oscillator,
for which the $\beta^4$ term confines the wave function. 
Iachello \cite{E5X5}, introduced two schematic potentials that  classify the structures in this region. The 
 "E(5)"   potential   does not depend on $\beta$ and $\gamma$ for $\beta<\beta_W$ where it jumps to $\infty$. The wave functions
 of the lowest states resemble the ones in the region of large values of $-d\lesssim-5$, where the potential weakly depends on $\gamma$.
 The "X(5)" potential has a term $\propto \gamma^2$ added to the E(5) potential. The corresponding wave functions resemble the ones in
the region $-5\lesssim d\lesssim1$, where the potential prefers the prolate shape. 
It has become  popular to use the acronyms E(5) and X(5) for these structures.

\begin{figure}[th]
%\centerline{\psfig{file=gcm_102pd_spec,width=\linewidth}}
\centerline{\psfig{file=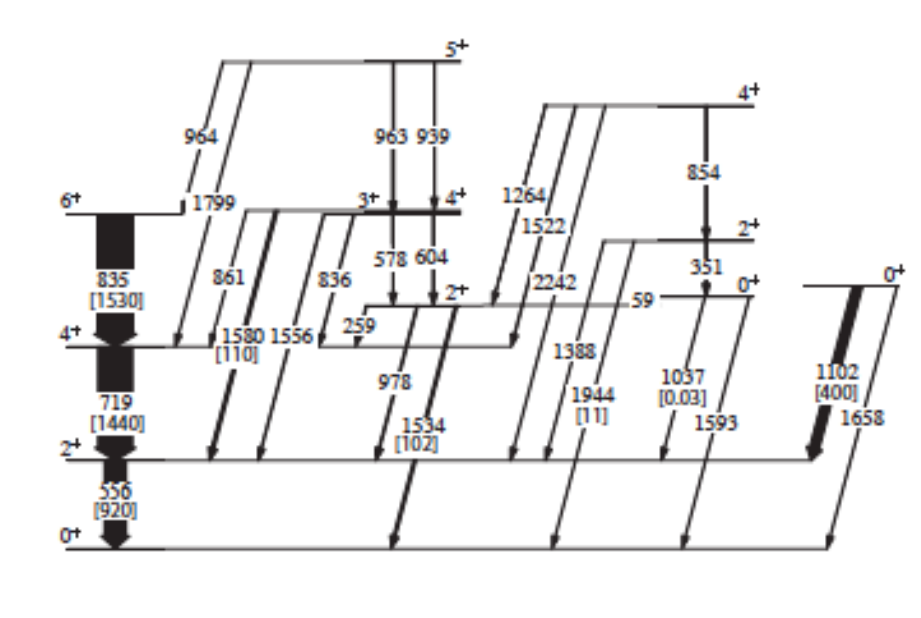,width=\linewidth}}
\caption{Experimental level scheme and $B(E2)$ strengths  for $^{102}$Pd. 
The number in parenthesis under the transition energies (keV) are the $B(E2)$ values ($e^2$ fm$^4$) for the transitions. 
Data from \cite{Zamfir02}. Preparation of the figure by A.D. Ayangeakaa is acknowledged. 
\label{fig:102PdSpecExp}}
\end{figure}

\begin{figure}[th]
\centerline{\psfig{file=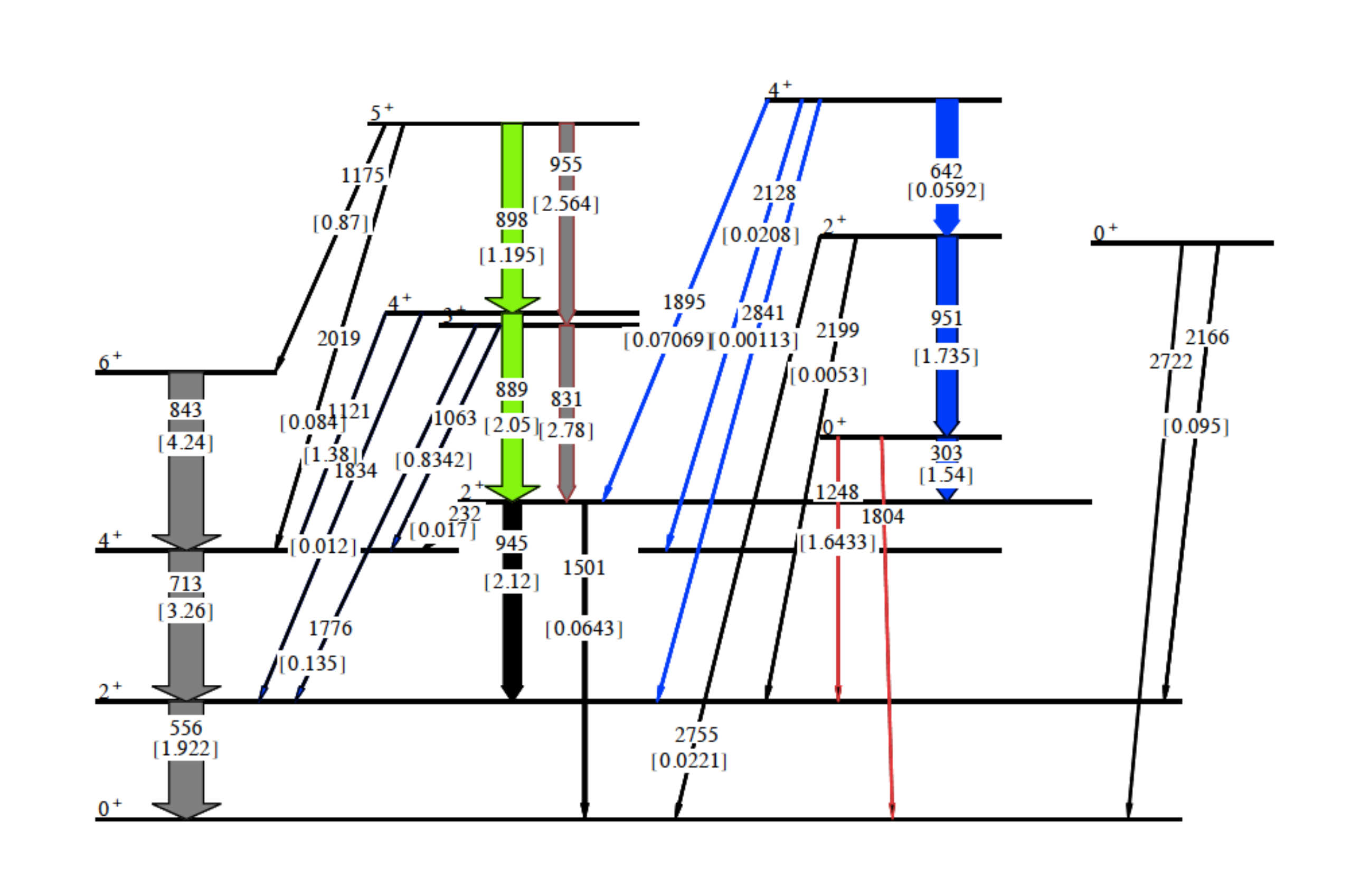,width=\linewidth}}
\caption{GCM predictions\cite{Zamfir02} of the energies and $B(E2)$ strengths for $^{102}$Pd  
 normalized to the experimental $E(2^+_1)$ and and $B(E2;2^+_1\rightarrow0^+_1)$
values. The number in parenthesis under the transition energies (keV) are the $B(E2)$ values ($e^2$ fm$^4$) for the transitions. 
%Observed levels with no clear theoretical
%counterpart or calculated levels with no clear experimental counterpart (see text) are indicated with dashed lines. 
 Preparation of the figure by W. Li is acknowledged.
\label{fig:102PdACM}}
\end{figure}

Fig. \ref{fig:102PdSpecExp} shows $^{102}$Pd as an example for an E(5) nucleus  \cite{Zamfir02}. 
Fitting the experimental energies and E2 transition probabilities provides  the GCM parameters $d=-43$ and
$\hbar^2S=24\times10^{-22}=56\times10^{42}(\mathrm {MeV}s)^{-2}\hbar^2$, which corresponds to $S/S_>=0.9$. The GCM
predictions are shown in Fig. \ref{fig:102PdACM}. The agreement with the data is 
characteristic for  GCM phenomenology.  The GCM predictions are similar to the ones of the E(5) model (e. g. Fig. 8  of Ref. \cite{Zamfir02}). 
An example for  the X(5) structure is $^{152}$Sm \cite{Casten01}. The GCM parameters are $d=-1.75$ and 
$\hbar^2S=71\times10^{-26}=18\times10^{39}(\mathrm {MeV}s)^{-2}\hbar^2$, which corresponds to $S/S_>=0.25$.
 The ratios of the energies and $B(E2)$ values from the X(5) model and from the GCM turn out to be  similar  as well (c. f. Ref. \cite{CaprioThesis}).
The reason of the similarity of the GCM fits and the schematic models is that the wave functions of the considered 
few lowest states do not resolve the details of the potential, because the wavelength is larger. 
Caprio  \cite{ Caprio2par}
presents figures of the ratios $E(I)/E(2^+_1)$ and $B(E2;I\rightarrow I')/B(E2;2^+_1\rightarrow0^+_1)$, which allows one to extract the 
GCM parameters, and discusses fitting strategies.   A GCM analysis of the $N=90$ region can be found in his thesis \cite{CaprioThesis}, 
where he also analyses the difference between the GCM and the schematic models.

\subsection{Triaxiality}\label{sec:triaxiality}

Strong  deviations from axial shape have been in the focus of recent research.  
The two-parameter version of the GCM encompasses the range between the axial regime with a stiff $\gamma$-vibration and
  Wilet-Jean (WJ) limit of a $\gamma$-independent potential.
To account for a stabilization of the triaxial shape, a term with a minimum at $\gamma=\pi/6$  has been added to the potential (\ref{eqn-VGCM2}), where
the  authors used different functions $f(\beta,\gamma)$ \cite{Bonsatos,RW,CaprioTriax}.  

Caprio \cite{CaprioTriax} studied the consequences of the gradual stabilization of 
 the triaxial shape in a systematic way for a frozen deformation $\beta=\beta_0$ by diagonalizing the Hamiltonian
 \begin{equation}\label{GCM3}
 H_{GCM3}=\Lambdahat^2+\chi\left[1-\cos3\gamma+\xi\cos^2 3\gamma\right].
  \end{equation}
As already noticed by Zamfir and Casten \cite{ZC91}, the order of the two signatures of the quasi-gamma band (2$^+_2$, 3$^+_1$, 4$^+_2$, 5$^+_1$, 6$^+_2$, ..) provides 
a clear spectral significance for the type of triaxiality.    The order is easily understood considering the two limiting cases. The $\gamma$-independent 
potential (WJ) has SO(5) symmetry. The states 2$^+_2$, (3$^+_1$, 4$^+_2$), (5$^+_1$, 6$^+_2$), ... belong to multiplets with SO(5) seniority v=1, 2, 3, ... , where
the energy is v(v+3)/4 in units of the $E(2^+_1)$.
That is, the even-$I$ states are below the average of their odd-$I$ neighbors. When the triaxiality parameter  is fixed at $\gamma=\pi/6$ (TR),
the ratio of the three moments of inertia is 1/1/4 (irrotational flow). The energies for this triaxial rotor (TR) 
are given by $I(I+1)/12+\left[2I(n+1/2)-n^2\right]/4$ in units of the $E(2^+_1)$,
where $n$ is the number of wobbling excitation and $I$ is even (odd) when $n$ is even (odd). That is, the odd-$I$ states are below the 
average of their even-$I$ neighbors. In the case of an axial rotor (AR), the two signatures of the $\gamma$ band are degenerate.

 \begin{figure}[th]
\centerline{\psfig{file=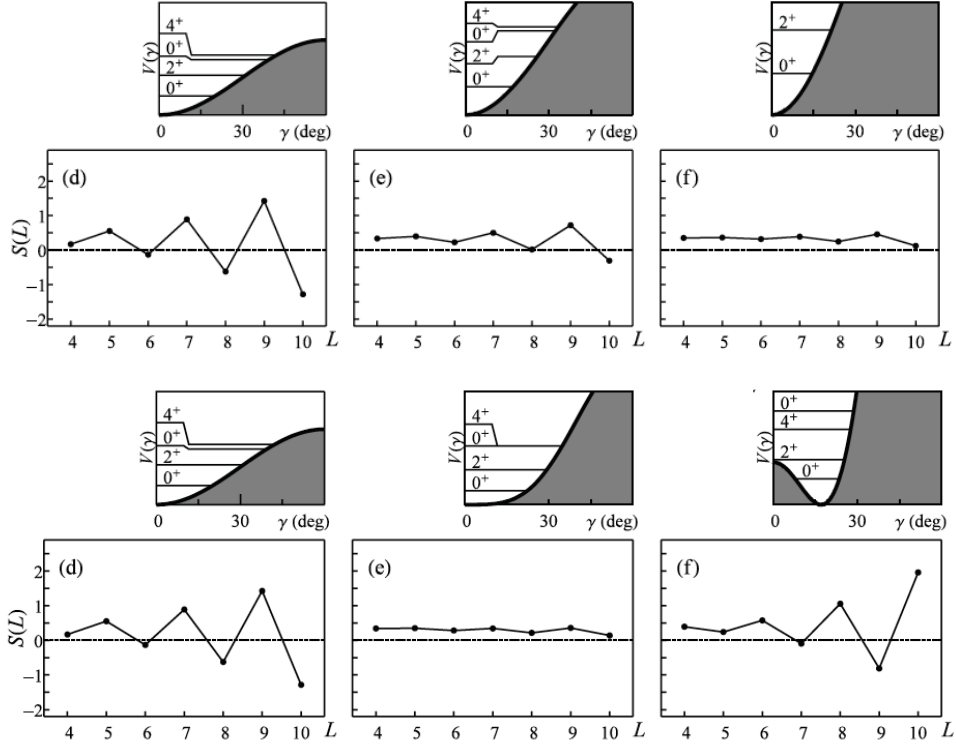,width=15cm}}
\caption{Staggering of level energies within the quasi-$\gamma$ band, as measured by the staggering parameter  $S(L)$ calculated
for the  Hamiltonian (\ref{GCM3}) with $\xi = 0$, for (d) $\chi = 50$, (e) $\chi = 100$, and (f) $\chi = 200$. in the upper
part and for (d) $\xi= 0$ with $\chi = 50$, (e) $\xi = 0.5$ with $\chi = 100$, and (f) $\xi = 0.8$ with $\chi = 500$ in the lower part.
 The potential $V (\gamma )$ is shown in the inset, with the ground, quasi-$\gamma$, and quasi-$\gamma\gamma$ band head
energies indicated.
 Reproduced and adapted  from Ref. \cite{Caprio2par}.}
\label{fig:StaggeringTh}
\end{figure}

 The staggering parameter
\begin{equation}
S(I)=(E(I)-2E(I-1)+E(I-2))/ E(2^+_1),
\end{equation}
  calculated from the energies $E(I)$ of the quasi-gamma band, is used to characterize the triaxiality \cite{ZC91}.  The limiting models give for,
 respectively   $I$ even/odd,
  \begin{eqnarray}\label{SImod}
 S_{WJ}(I)=\frac{1}{8}\mp\frac{1}{4}\left(I+\frac{9}{2}\right), ~~S_{TR}(I)=\frac{1}{6}\pm\left(I-\frac{5}{2}\right), ~~~S_{AR}=\frac{1}{3}.
 \end{eqnarray} 
 Another spectral significance is the ratio $E(2^+_2)/E(4^+_1)$ which is $>1$, 1, $<1$ for the AR, WJ, TR, limits respectively. 
 Caprio \cite{CaprioTriax} studied in detail the  transition between the regimes in the framework of the model (\ref{GCM3}) and provided 
 the figures and tables of the ratios of the energies and of $B(E2)$ values   that allows one to determine the model parameters.
 Fig. \ref{fig:StaggeringTh} illustrates in the upper part the development from the AR to the WJ regime and in the lower part 
 the development from the WJ to the TR regime. It is noted that small staggering is not necessarily an indication for the common case of small triaxiality. 
 It also appears when the staggering phase changes sign at the transition from the WJ to the TR regime. The low ratio $E(2^+_2)/E(4^+_1)\leq 1$
 can be used to discriminate this case from axiality.

 \begin{figure}[th]
\begin{center}
\psfig{file=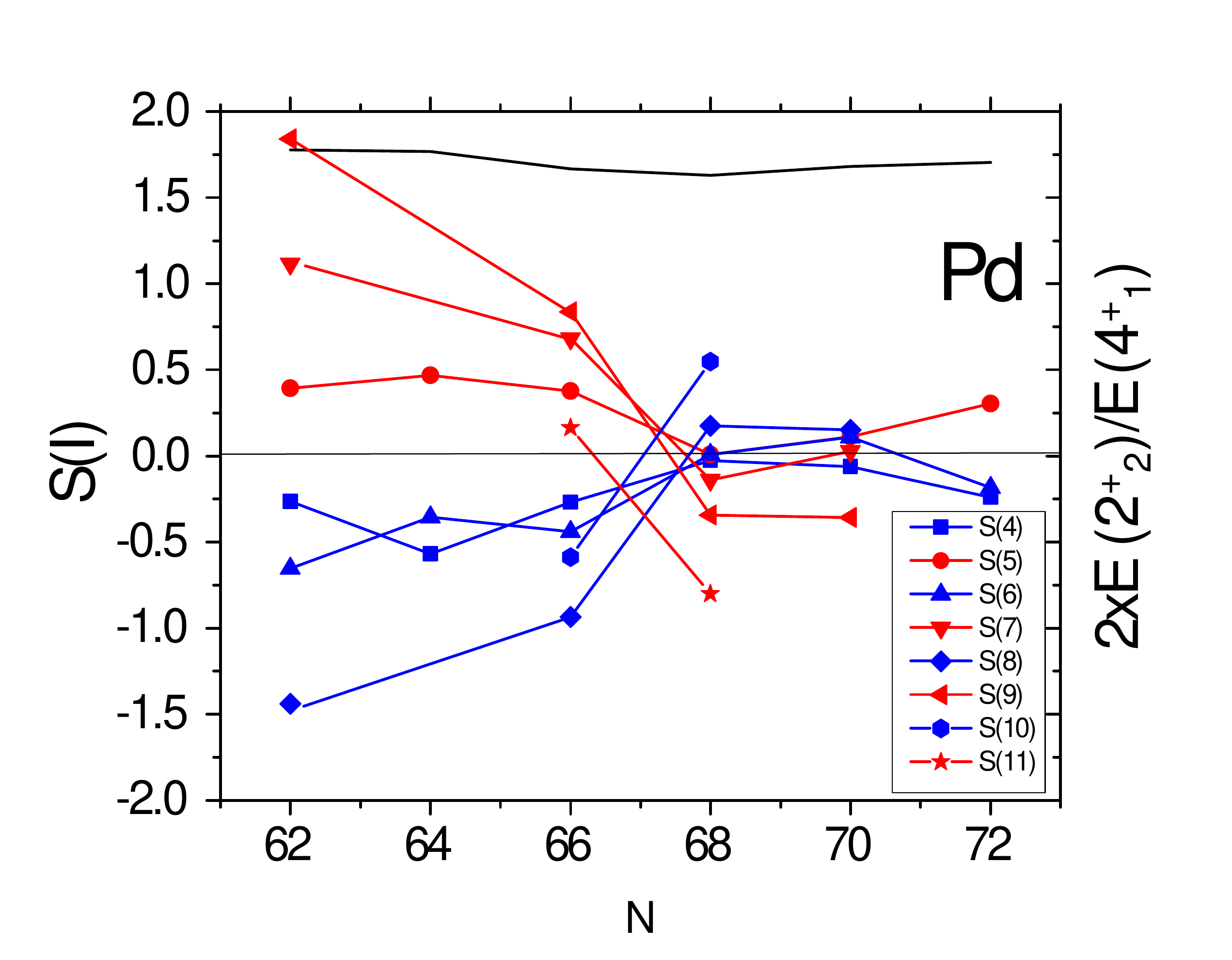,width=10cm}
\caption{\label{fig:StaggeringPd} Experimental staggering parameter $S(J )$ for the Pd isotopes. The black line 
shows the ratio $E(2^+_2)/E(4^+_1)$. Data from ENSDF \cite{ensdf} and Ref. \cite{Luo13}}.
\end{center}
\end{figure}
\begin{figure}[h]
\begin{center}
\psfig{file=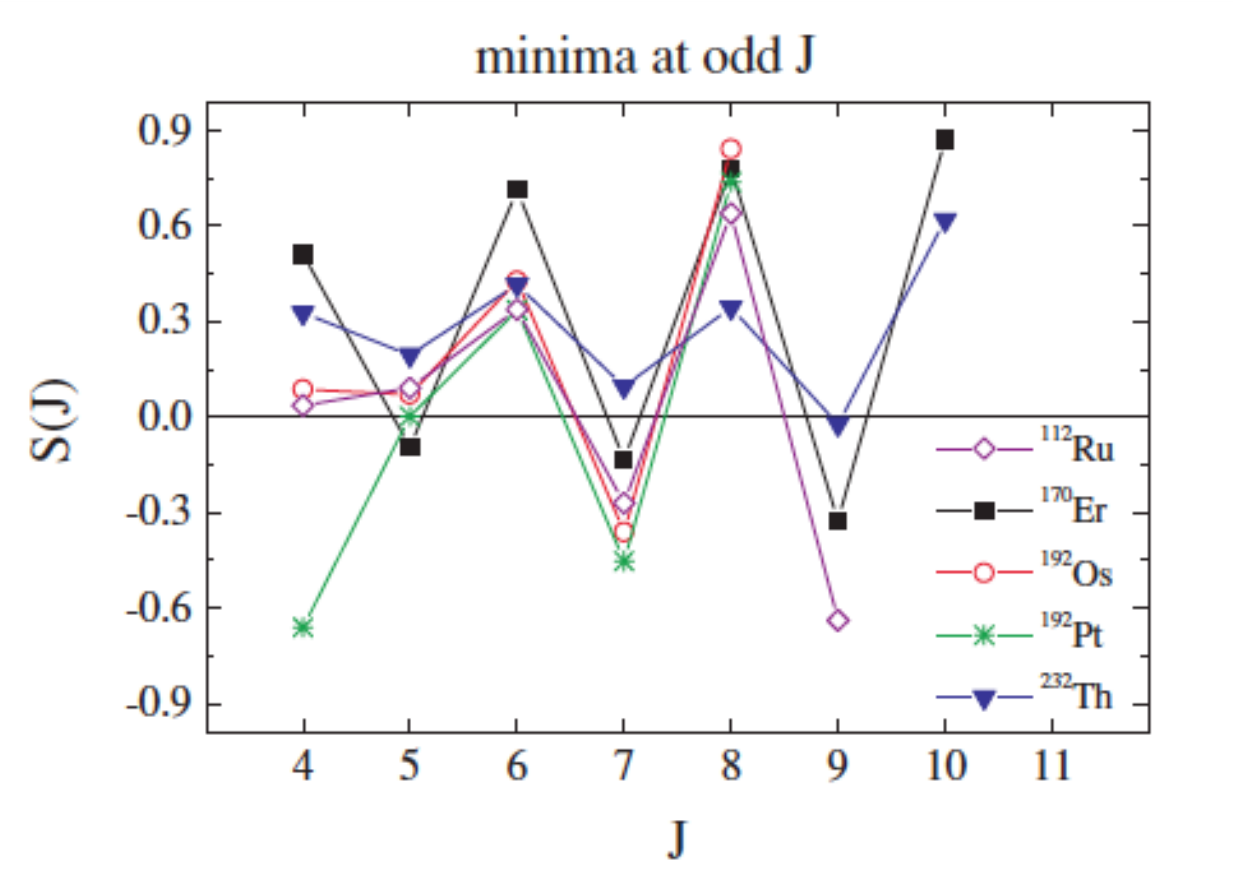,width=10cm}
\caption{\label{fig:StaggeringOdd} Experimental staggering parameter $S(J )$ for nuclides that shows the staggering associated
with triaxial shapes. Reproduced from Ref. \cite{McCutchan07}.}
\end{center}
\end{figure}

As an example, Fig. \ref{fig:StaggeringPd} demonstrates the change of triaxiality along the Pd isotope chain. 
The ratio  $E(2^+_2)/E(4^+_1)$ stays around 0.85, with a very shallow dip around N=68, which 
 indicates strong deviations from axiality. Below $N=66$ one sees the 
$\gamma$-soft pattern: even $I$ below odd $I$. For $N=68$ and 70, the triaxial rotor pattern emerges  for $S(I>5)$: odd $I$
below even $I$. There is a vague indication that the   $\gamma$-soft pattern returns for $N=72$.  
 The  Ru chain shows a similar trend as the Pd chain, where  the amplitude of the 
 staggering is larger: $S(8)-S(9)=1.2$ in $^{112}$Ru and $S(8)-S(9)=0.6$ in $^{114}$ Pd. 
 McCutchan {\it et al.} \cite{McCutchan07} studied the systematics of the staggering parameter. 
  Fig. \ref{fig:StaggeringOdd}  from their work collects nuclei with the odd-$I$-low staggering of a stabilized triaxial shape. 
  The other investigated nuclides either show
  the $S$-positive pattern of axial shape or the  even-$I$-low pattern of $\gamma$-soft nuclei. Toh {\it et al.} 
  \cite{Toh13} reported odd-$I$-low staggering for $^{76}$Ge, where the neighboring nuclides show
  the even-$I$-low pattern. However this cannot be taken as "Evidence for rigid triaxial deformation at low energy"
  as claimed by the authors, because the observed value of $S(8)-S(9)=0.5$ is an order of magnitude smaller than
  the rigid rotor value of 12 given by Eq. (\ref{SImod}).  The nucleus seems to be located  in-between the cases e) and f) shown 
  in the lowest panel of Fig. \ref{fig:StaggeringTh}. That is, the zero-point energy is substantially larger than the depth of the potential minimum at 
  triaxial shape, which is far from a rigid-triaxial regime. The authors carried out Shell Model calculations, which well reproduce both 
  experimental energies and $B(E2)$ values (amplitude of the staggering in particular). This is expected because the Shell Model
  takes the dynamics of the quadrupole degree of freedom into account.
    
  \section{Adiabatic Time-Dependent Mean Field approach}\label{sec:ATDMF}
  The  Adiabatic Time-Dependent Mean FieId (ATDMF) approach is the standard method for deriving the parameters of the BH from 
  the Fermionic underpinning.   Baranger and Kumar introduced it in  their pioneering work \cite{BK68}.  ADTMF is very well
  exposed in the reviews by Prochniak and Rohozinski \cite{prochniak2009} and Nik\v{s}i\'{c}, Vretenar and Ring \cite{NVR11}.
  Here we only sketch few steps that are essential for
  the discussion to be followed. 
  
  First, the potential is obtained by the mean field approach of choice, which  is constrained to  provide a given
  value of the  expectation value of the microscopic mass quadruple moments $q_0=\left< Q_0\right>$ and $q_2=\left<Q_2\right>$.
  The deformation parameters $q_0$ and $q_2$ and the three Euler angles, which specify the orientation of the quadrupole shape,
  are the dynamic variables of the BH.   To keep contact with phenomenology one often introduces the standard dimension-less
  deformation parameters $\beta$ and $\gamma$ by the relation
  \begin{equation}\label{BetaGammaMicro}
  q_0=c\beta\cos\gamma,~~~q_2=c\beta\sin\gamma,~~~c=\sqrt{\frac{5}{\pi}}A\frac{3}{5}\left(r_0A^{1/3}\right)^2,~~~r_0=1.2fm,
    \end{equation}
   which assumes the liquid drop relation between the deformation parameters and the quadrupole moments.
 Second, a classical BH is derived by time dependent perturbation theory. The adiabatic approximation is used, which 
 means that only terms quadratic in the time derivative of the expectation values $\left<Q_\mu\right>$ are kept. 
 \begin{eqnarray}\label{BHclass}
 H_{class}=T_{vib}+T_{rot}+V(\beta,\gamma),\\
 T_{vib}=\frac{1}{2}B_{\beta\beta}\dot\beta^2+\frac{1}{2}\beta^2B_{\gamma\gamma}\dot\gamma^2+B_{\beta\gamma}\beta\dot\beta\dot\gamma,\\
 T_{rot}=\frac{1}{2}\sum_{i=1,2,3}{\cal J}_i\omega_1^2, ~~~~{\cal J}_i=4B_i\beta^2\sin^2(\gamma -\frac{2}{3} \pi i),
   \end{eqnarray}
   where $\omega_i$ are the angular velocities with respect to the three body fixed axes. 
 The classical BH corresponding to the GCM Hamiltonian (\ref{eqn-HGCM2}) is obtained by setting $B_{\beta\beta}=B_{\gamma\gamma}=B_i=\sqrt{5}/2B_2$
 and $B_{\beta\gamma}=0$. The microscopically calculated mass parameters are far from this drastic simplification (see below), which is a severe limitation
 of the GCM phenomenology.  
 
 Third, the classical BH (\ref{BHclass}) is quantized. The procedure is involved because   the mass parameters depend on  the deformation.
  It is described 
 in the reviews \cite{prochniak2009,NVR11}, where the resulting complicated expression  for the quantal BH is quoted and the numerical 
 methods for the solution of the eigenvalue problem are discussed.  
 The calculation of the potential and the mass parameters depends on the mean field approach of choice. 
 It has recently become customary to refer to the various variants as 5DBH (5 Dimensional Bohr Hamiltonian).
 
 \subsection{Pairing plus Quadrupole Model}\label{sec:5DBH-PQQ}
 
  The generic version of the microscopic BH \cite{BK68}, 5DBH-PQQ, is derived from the Pairing plus Quadrupole-Quadrupole (PQQ) Hamiltonian, 
  \begin{equation}\label{HPQQ}
  H_{PQQ}=h_{sph}-GP^\dagger P-\frac{\kappa}{2}\sum_\mu Q^\dagger_\mu Q_\mu,
  \end{equation} which 
 combines a spherical term $h_{sph}$ constructed from adjusted spherical single particle energies  with a Monopole Pairing interaction for the short-range correlations and
 a Quadrupole-Quadrupole interaction for the long-range correlations.  It generates the mean field Hamiltonian
 \begin{equation}\label{hPQQ}
 h_{PQQ}=h_{sph}+\Delta\left(P^\dagger+P\right)+\kappa q_0Q_0+\kappa q_2\left(Q_2+Q_{-2}\right)-\lambda N.
 \end{equation}
 This is the BCS Hamiltonian for a deformed Nilsson-type potential, which has the standard BCS ground state $\vert\rangle=\vert \Delta, \lambda,q_0,q_2\rangle$. 
  (Protons and neutrons are not explicitly distinguished for simplicity.). The term $-\lambda N$ assures the correct expectation value of the particle number by the equation  $\left<N\right>=N$. 
 The pair field is determined by the selfconsistency equation $\Delta=G\left<P\right>$. 
 The pair field $\Delta(q_0,q_2)$ and the chemical potential $\lambda(q_0,q_2)$ are functions of the deformation parameters.
 
  The mass parameters take the Inglis-Belyayev (IB) form with $\mu=0,2$
 \begin{eqnarray}
 {\cal M}_{(n),\mu\nu}=\sum_{kl}\frac{\left<\vert Q_\mu\vert kl\right>\left<kl\vert Q_\nu\vert \right>}{\left(E_k+E_l\right)^n}\label{Mn}\\
 B_{\mu\nu}=\frac{(\hbar\kappa)^2}{2}{\cal M}_{(3),\mu\nu}+P_{\mu\nu},~~~
 {\cal J}_i=\sum_{kl}\frac{\left<\vert J_i\vert kl\right>\left<kl\vert J_i\vert \right>}{\left(E_k+E_l\right)},\label{Bcranking}
    \end{eqnarray}
 where $\vert\rangle$ and $\vert kl\rangle$ stand for the zero- and two-quasiparticle states, and $E_i$ are the quasiparticle energies. The term $P_{\mu\nu}$
 takes the deformation dependence of $\Delta$ and $\lambda$ into account.  The somewhat complicated expression is given in Ref. \cite{BK68}.
 The mass parameters for the dimension-less deformation variables $\beta$ and $\gamma$ are obtained by transforming the matrix $B_{\mu\nu}$ 
 as $c{\cal R}(\gamma)^TBc{\cal R}(\gamma)$, where $c{\cal R}(\gamma)$ is a rotation by $\gamma$ and a multiplication by the scale factor $c$.  
 
 The PQQ model takes two shells into account. The pairing strength is adjusted to the experimental even-odd mass differences. The quadrupole coupling
 constant $\kappa$ is treated as a parameter  that is adjusted to reproduce the experimental spectra.
 The IB mass parameters turn out to be to small. A common scaling factor$F_B\sim2$ to 3 is multiplied to 
 obtain the experimental energy scale. For the $E2$ matrix elements a polarization charge of $kZ/A$  with $k\sim 1.6$ is used. The two parameters set the
  scales for the energy and charge quadrupole moments. 
 Compared to the GCM and IBM phenomenology, which use the same free scales, the 5DBH-PPQ model has only one parameter, $\kappa$, which  changes 
 smoothly $\propto A^{-1.4}$ through  regions of major re-structuring. The nature of collective excitations is determined
 by the shell structure of the nucleonic orbitals. In case of the PQQ, it is encoded in the spherical single particle levels, which are input as well, though determined from
 data different from the collective excitations.

  \begin{figure}[t]
\begin{center}
\psfig{file=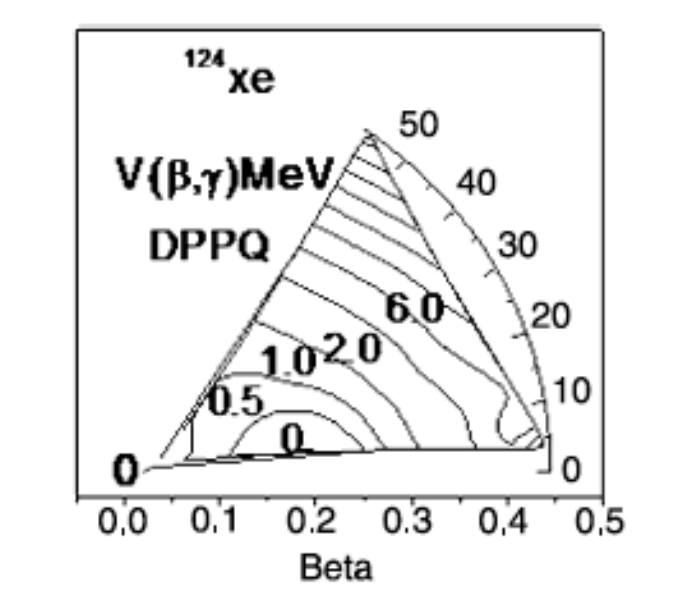,width=4.27cm}
\psfig{file=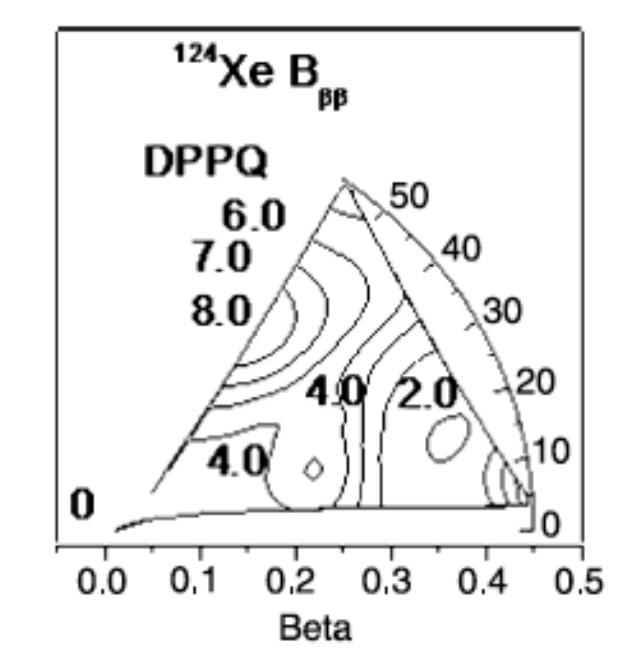,width=3.6cm}
\psfig{file=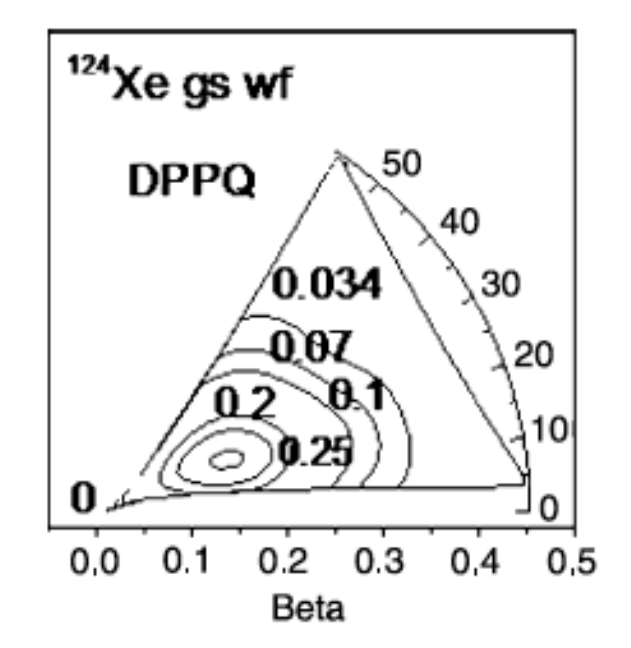,width=3.71cm}
\caption{\label{fig:VBbbWf124Xe} The potential $V$ (left), the mass coefficient $B_{\beta\beta}$ (middle, arbitrary units)
and the ground state wave function (right) calculated by means of the 5DBH-PPQ for $^{124}$Xe. Reproduced with permission from Ref. \cite{Gupta14}}
\end{center}
\end{figure}

 %Kumar and Gupta applied the 5DBH-PPQ, which they call Dynanic Pairing Plus Quadruple (DPPQ) model, to the following nuclei  
 %$^{182-186}_{74}$W, $^{186-192}_{76}$Os,  $^{192}_{78}$Pt Ref. \cite{KB68},
 %$^{166,168}_{72}$Hf  Ref. \cite{Gupta12}. They carried out a 
 %detailed study of phase transition region around $N=90$:
  %$^{150,152}_{62}$Sm Ref. \cite{K74},
  %$^{142-154}_{62}$Sm Ref. \cite{Gupta83},
  %$^{150}_{62}$Sm Ref. \cite{Gupta10}
  %$^{154,156}_{64}$Gd Ref. \cite{Gupta77},
  %$^{154,156}_{66}$Dy Ref. \cite{Gupta89}
  %$^{144-150_{60}}$Nd Ref. \cite{Gupta88},
  %and further studied: 
   %$^{130-136}_{58}$Ce Ref. \cite{GK12},
   %$^{122-134}_{56}$Ba Ref. \cite{KG01}, the   
   %$N=66$ isotones $_{42}$Mo-$_{58}$Ce Ref. \cite{GK02}, and     
   %$^{124}$Xe Ref. \cite{Gupta14}. In addition to the energies and E2 matrix elements, they
   %calculated magnetic moments and spectroscopic factors and E0 moments.   

      Kumar, Baranger, and Gupta applied the 5DBH-PQQ, which they call Dynamic Pairing Plus Quadruple (DPPQ) model, to  nuclei  
      with ($Z, A$)=
 (74,182-186), (76,186-192),  (78,192) in Ref. \cite{KB68},
 (72,166-168) in  Ref. \cite{Gupta12},
  (62,150-152) in Ref. \cite{K74},
  (62,142-154) in Ref. \cite{Gupta83},
  (62,150) in Ref. \cite{Gupta10}
  (64,154-156) in Ref. \cite{Gupta77},
  (66,154-156) in Ref. \cite{Gupta89},
  (60,144-150) in Ref. \cite{Gupta88}, 
   (58,130-136) in Ref. \cite{GK12},
   (56,122-134) in Ref. \cite{KG01}, 
   (54,124) in Ref. \cite{Gupta14}, 
   and the   $N=66$ isotones with $Z=42-58$ in Ref. \cite{GK02}.     
    In addition to the energies and E2 matrix elements, they
   calculated magnetic moments, spectroscopic factors, and E0 moments.   
  They investigated the phase transition region around $N=90$ in considerable detail, subsequently tuning
  the model parameters. More recent work extensively compares the  5DBH-PQQ with the IBM phenomenology.

As an example, consider  $^{124}$Xe.  5DBH-PQQ well reproduces both energies and $B(E2)$ values.  Fig. \ref{fig:VBbbWf124Xe} illustrates  the calculations. 
The potential has a minimum at axial shape around $\beta$=0.24. The mass coefficient $B_{\beta\beta}$ is  far from being constant, as assumed in the GCM phenomenology.  
 It increases toward 
small $\beta$ on the prolate side and becomes large on the oblate side. The wave function tends to concentrate in regions of large mass like in regions of low potential. 
This has the consequence that the wave function has its maximum around  $\beta=0.16$ and does not change much in $\gamma$-direction. That is, it has the character 
of the WJ model. The calculated staggering of the quasi-$\gamma$ band is of the even-$I$-low type in agreement with the experiment, although the model overestimates the
experimental staggering parameter: $S(5)$ = 208 keV and $S(6)$= -92 keV  to be compared with the calculated values  498, -475 keV, respectively.  
The ratio $B(E2, 3^+_\gamma \rightarrow 2^+_g)/B(E2, 3^+_\gamma \rightarrow 2^+_\gamma)$ is small (0.024 experimental and 0.064 calculated) whereas 
the ratio $B(E2, 3^+_\gamma \rightarrow 4^+_g)/B(E2, 3^+_\gamma \rightarrow 2^+_\gamma)$ is large (0.26 experimental and 0.19 calculated), which reflects the seniority
selection rules of the WJ limit corresponding to 0 and 0.4, respectively. The calculated energy of the $\gamma$ band head $E(^+_\gamma)$ = 1097 keV is larger than
the experimental value of 847 keV and the calculated  $E(0^+_2)$ = 1099 keV is lower than the experimental value of 1269 keV. The example has  the 
typical accuracy the   5DBH-PQQ calculations. It demonstrates that the just looking on the calculated potentials for guessing the structure of the low-lying collective excitations may be risky, 
because the deformation dependence of the mass coefficients may cause substantial modifications of the wave functions. Conversely,   a potential derived by fitting data in the framework 
of the GCM, which assumes a simple deformation dependence irrotational flow for the mass coefficients,  may substantially differ from the microscopic potential calculated by 
the mean field approach.   
\begin{figure}[t]
\begin{center}
\psfig{file=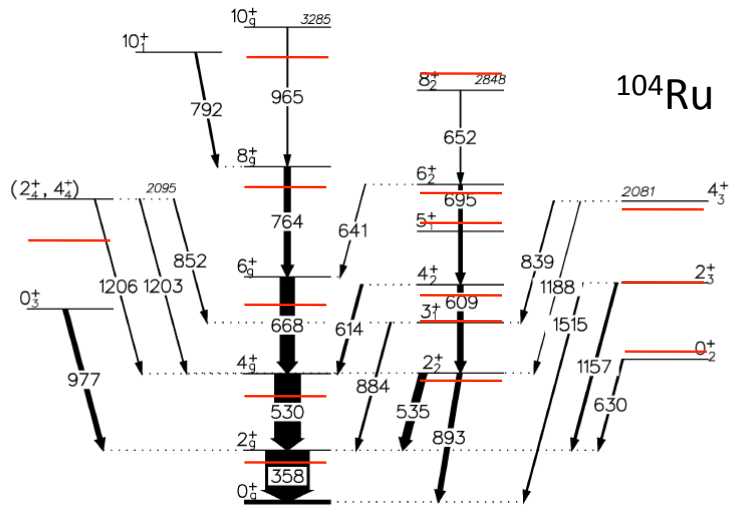,width=\linewidth}
\caption{\label{fig:104RuS} The experimental level scheme of $^{104}$Ru (black levels)  compared with the ones calculated by means of the 5DBH-SC model (red levels). 
  Reproduced from Ref. \cite{Srebrny06}}
\end{center}
\end{figure}   
 
\begin{figure}[t]
\begin{center}
\psfig{file=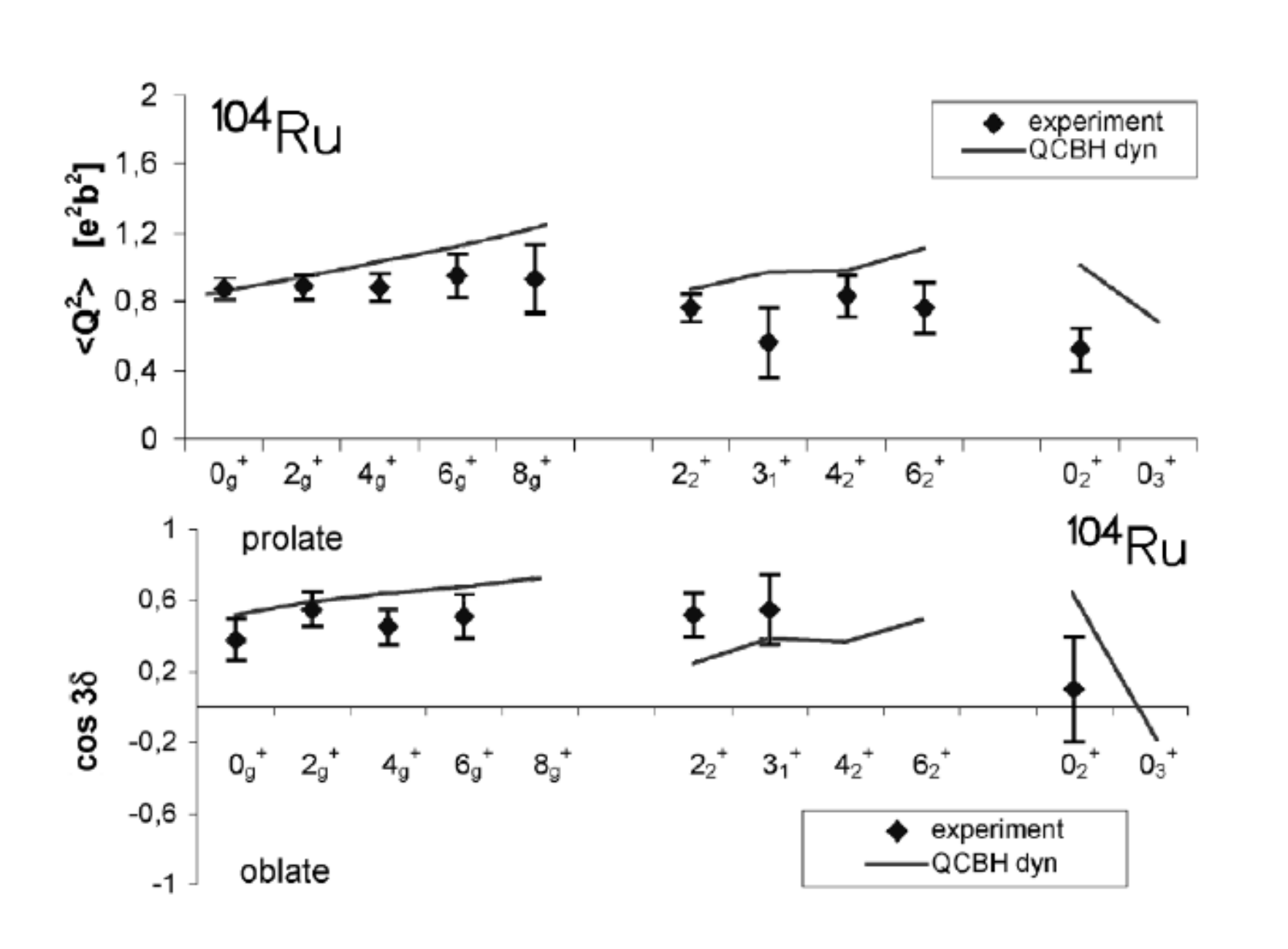,width=\linewidth}
\caption{\label{fig:104RuIn} The experimental invariants $\left<Q^2\right>=s^2\left<\beta^2\right>$ and 
$\cos 3\gamma=\left<\beta^3 \cos 3\gamma\right>/\left<\beta^2\right>^{3/2}$ for 
 $^{104}$Ru compared with the ones calculated by means of the 5DBH-SC model (QCBH dyn). 
  Reproduced from Ref. \cite{SC11}.}
\end{center}
\end{figure}    

 \subsection{Pairing plus Quadrupole Model with Local Random Phase Approximation}\label{sec:5DBH-PQQ+LQRPA}
  Matsuo, Matsuyangi, Nakatsukasa {\it et al.} \cite{Matsuyanagi10,Matsuo14} removed the problem with the too dilute energy scale of the 5DBH-PQQ approach by adding a
  quadrupole pairing interaction to the PQQ Hamiltonian (\ref{HPQQ}).  Quadrupole pairing is known to increase  the 
  rotational moment of inertia by about 30\% when calculated by means of the self-consistent cranking model.  The pair field generated by the 
  quadrupole pairing adds a new term to the IB value, which  increases it to the Thouless-Valatin (TV) value. The Adiabatic Self-consistent  Collective Coordinate
  (ASCC) method developed by the authors allows them to calculate the TV correction to the IB values for the mass parameters $B_{\beta\beta}$ and $B_{\gamma\gamma}$ as well.
  In practice, the equations of the Quasiparticle Random Phase Approximation (QRPA) are solved for each point of the $\beta$-$\gamma$ grid (Local QRPA -LQRPA). 
  The mass parameters are derived from the frequency and oscillator length of the lowest two solutions in the standard way. 
  An increase of 20\% to 30\% of all mass parameters of the BH is found, which results in the  energy scale of the experiment. 
 
Carrying out the  LQRPA for the large number of  grid points is computational extensive. For this reason only nuclei with mass below 100 have been studied:
$(Z,~N)=(34,~34-38)$ in Ref. \cite{Hinohara10Se}, (24, 34-44) in Ref. \cite{Yoshida11Cr,Sato12Cr}, and (12, 18-22) in Ref. \cite{Hinohara11Mg}.  Without any free parameters, 
the agreement of the calculations with the experiment is comparable with the 5DBH-PQQ, which uses a scaling factor. The analysis in Ref. 
\cite{Sato12Cr,Hinohara11Mg} demonstrates that the two-parameter  BH phenomenology in Sect. \ref{sec:GCMphen} is not capable of adequately  accounting
for the microscopy based  wave functions.  In the case of 0$^+_2$ states, the deformation dependence of the mass parameters leads to 
structures in between the $\beta$-vibration and shape coexistence.

   \subsection{The Micro-Macro Model}\label{sec:mic-mac}\label{sec:5DBH-SC}
   
   Rohozi\'nski {\it et al.} \cite{Roho77} introduced the 5DBH-SC  version of the microscopic BH. The potential energy $V(\beta,\gamma)$ is obtained by means of Strutinski's 
   Shell Correction (SC) method, which combines the deformation energy of a droplet of nuclear matter with a shell correction  obtained from the single particle levels in a deformed 
   modified oscillator (Nilsson) potential. The SC method, which is now often called the Micro-Macro approach, is well tested for calculating deformation energies. It uses the spherical 
   single particle energies and the surface and Coulomb energy of the nuclear droplet as input, which are determined from data other then the low-energy collective excitations. 
   The scale of the E2 matrix elements is fixed by the Coulomb field of the droplet. 
   The mass parameters are calculated  by means of the IB expressions (\ref{Bcranking}), which are
    too small when the standard BCS monopole pairing is used. Pr\'ochniak {\it et al.} \cite{Proch99}  removed the problem by improving the treatment of the pair correlations.
   They construct a collective Hamiltonian for the  pair gap $\Delta$ and the  gauge angle $\Phi$ in analogy to the BH for the quadrupole degrees of freedom.
   For each point on the $\beta$-$\gamma$ grid, they calculate the pairing ground state wave function and determine the most probable value 
   of $\Delta$, which they the use in calculating the IB mass parameters.
   The most probable value of $\Delta$ is smaller than the BCS value and depends less on deformation, which brings the mass parameters to the right scale.
   With this modification the 5DBH-SC becomes parameter-free.  Pr\'ochniak  \cite{Proch05} demonstrated that including the direct coupling between the quadrupole and the pairing degrees of 
   freedom in the framework of a nine dimensional generalized BH provides essentially the same results as the approximation \cite{Proch99} of using the the most probable $\Delta$ values.
  The fact that the right scale of the mass parameters can be achieved either  by the dynamical treatment of the monopole pairing or by inclusion of the quadruple pairing (see 
  preceding Sec. \ref{sec:5DBH-PQQ+LQRPA}) raises the question about the consequences of combining both improvements of the BCS treatment.     
   
   The 5DBH+SC has been applied to  nuclides with $Z,~N)=$
  (52-62, 68-80) in Ref. \cite{Zajac99}, (46, 60-64), (44, 60-70) in Ref. \cite{Proch99}, and (44, 60) in Ref \cite{Srebrny06}. As an example, Fig. \ref{fig:104RuS} 
  demonstrates that the 5DBH+SC very well reproduces the spectrum of $^{104}$Ru without  fitting any parameters. The 5DBH+SC has been preferentially applied to nuclei for which
   the $E2$ matrix elements have been measured by means of Coulomb excitation, providing a  stringent test, which is well met by the model. In addition to direct comparison, the  authors 
   studied invariants of the quadrupole mode, which are derived by combining measured transition matrix elements. Srebrny and Cline describe the method in Ref. \cite{SC11}, where they discuss
   a number of examples and expose the history.  It exploits the sum rules
   \begin{eqnarray}
   \frac{s^2}{\sqrt{5}}\left<II\alpha\vert\beta^2\vert II\alpha\right>=\sum_{\mu\nu}\left<2\mu2\nu\vert00\right>\left<II\alpha\vert Q_\mu Q_\nu\vert II\alpha\right>~~~~~\nonumber\\
   =\sum_{\mu\nu I'M'\alpha'}\left<2\mu2\nu\vert00\right>\left<II\alpha\vert Q_\mu\vert I'M'\alpha'\right>\left<I'M'\alpha' Q_\nu\vert II\alpha\right>~~~~~\\
  -\frac{\sqrt{2}s^3}{\sqrt{35}}\left<II\alpha\vert\beta^3\cos 3\gamma\vert II\alpha\right>
  =\sum_{\mu\nu\kappa\lambda}\left<2\mu2\nu\vert2\kappa\right>\left<2\kappa2\lambda\vert00\right>,
   \left<II\alpha\vert Q_\mu Q_\nu Q_\lambda\vert II\alpha\right>\nonumber~~~~~\\
   =\sum_{\mu\nu\kappa\lambda I'M'\alpha'I''M''\alpha''}\left<2\mu2\nu\vert2\kappa\right>\left<2\kappa2\lambda\vert00\right>~~~~~\nonumber\\
   \times\left<II\alpha\vert Q_\mu\vert I'M'\alpha'\right>\left<I'M'\alpha'\vert Q_\nu \vert I''M''\alpha''\right>\left<I''M''\alpha''\vert Q_\lambda\vert II\alpha\right>,~~~~~
      \end{eqnarray} 
   which are evaluated with the experimental matrix elements. The theoretical expectation values are directly calculated by integrating the wave functions.
   Fig. \ref{fig:104RuIn} exemplifies  the accuracy of the model for $^{104}$Ru. 
   The calculations overestimate the increase of the deformation with $I$ within  of the ground band. The nucleus is $\gamma$-soft with a preference of prolate shape,
   which is consistent with its even-$I$-low staggering of the $\gamma$ band. 

\subsection{Mean Fields on a Microscopic Basis}
More recent versions of the microscopic BH are based on mean fields derived from  the Skyrme Energy Density Functional (SK-EDF),
the Relativistic Mean Field approach (RMF), and the Hartree-Fock-Bogolybov method applied to the Gogny effective interaction (GI). Compared 
to the PQQ and SC versions of 5DBH, the spherical single particle energies are no longer phenomenological input (either direct or as the parameters of the Nilsson
potential),  but derived from a more fundamental layer. However this does not necessarily imply that they
reproduce the shell structure more accurately than the SC or PQQ approaches. The detailed $Z-,~N$-dependence of the binding energies  is excellently reproduced by the 
Finite Range Droplet Model (FRDM) by M\"oller and collaborators, who base the SC approach on the folded 
Yukawa potential (c. f. Ref. \cite{Moller06} and references therein). The spectral properties
obtained by the 5DBH are very sensitive to the shell structure, and therefore to the accuracy the mean field reproduces it.  In view of this, an application of the 5DBH-SC
method to the FRDM  promises a relatively simple and computational inexpensive way  of predicting the collective quadrupole excitations of even-even 
nuclei.
  
Deriving the 5DBH from the modern mean field approaches requires several sophistications, which are well exposed in the
review articles \cite{prochniak2009,NVR11}. 
The definition of the collective coordinates is less direct than for the PQQ and SC versions. The collective coordinates
 $c\alpha_\mu(\lambda_{2\nu})=\left<mf,\lambda_{2\nu}\vert Q_\mu\vert mf,\lambda_{2\nu}\right>$ 
 are implicit functions of the Lagrange parameters of the constraints $-\lambda_{2\nu} Q_\nu$ that 
 are  used  to generate the manifold of mean field states $\vert mf,\lambda_{2\nu} \rangle$. This leads to a modification of the IB mass parameters of the form 
 $2B_{\mu\nu}=\hbar^2\left({\cal M}_{(1)} ^{-1}{\cal M}_{(3)} {\cal M}_{(1)} ^{-1}\right)_{\mu\nu}$.
 The interaction induces a time-odd terms in the mean field, which generate corrections to the IB values of the mass parameters increasing them
 to TV values. Since their evaluations is difficult (see Sec. \ref{sec:5DBH-PQQ+LQRPA}),  they are usually neglected or only partially included.
 Further, there are zero point energy corrections (ZPE), which can be derived in different ways. 
 Usually the expressions obtained from the Generator Coordinate Method with the Gaussian Overlap approximation
 (GCM+GOA, see below) are used. The ZPE consist a rotational part     $\Delta V_{rot}=\sum_{\mu=-2,-1,1}{\cal M}_{(2),\mu\mu}/4{\cal M}_{(3),\mu\mu}$ and a vibrational part  
 $\Delta V_{vib}=Tr\left({\cal M}_{(2)}{\cal M}_{(3)}^{-1}\right)/4$.     
 
Pr\'ochniak {\it et al.} \cite{Proch04} introduced the 5DBH+SK, which is based on the Skyrme EDF. They use the IB mass parameters, no ZPE, and either a monopole pair interaction or 
a zero-range pair interaction. The mass parameters turn out to be too small, which is corrected by scaling the energies by a factor of 1.2 -1.3.   They studied the nuclides with $(Z,~N)$=
(40-42,62), (44,66), (56, 64), (54-56,70) in Ref.  \cite{Proch04}, (92-94,146), (96,150-152), 98,152-154) in Ref. \cite{Proch08},  (48, 52-58)
in Ref. \cite{Proch12},  (42, 42-58) in Ref. \cite{Proch10}, (42, 54-58) in Ref. \cite{Wrzosek11}, (42, 58) in Ref. \cite{Wrzosek12}, and (36, 36-40) in Ref. \cite{Proch09}. 
Like the 5DBH+SC, the 5DBH+SK has been preferentially applied to nuclei for which the $E2$ matrix elements have been measured by means of Coulomb excitation.

The Bejing-Munich-Zagreb collaboration worked out 5DBH+RMF, which is based on the Relativistic Mean Field approach. It is well exposed in the review \cite{NVR11}.
 It uses IB mass coefficients and  ZPE corrections. The pairing 
is taken in mean field approximation  for a density-dependent $\delta$-interaction or a separable in momentum space interaction. In most cast cases, the IB mass coefficients give  a too 
diluted spectrum, which is corrected by a scaling factor. An efficient method to calculate the TV contributions has been suggested in Ref. \cite{Li12}, however not yet implemented.
They studied the nuclides with $(Z,~N)$= (94, 156), (68, 98) in Ref. \cite{Li10}, (60, 84-94), (62, 88-90), (64, 88-90) in Ref. \cite{Li09}, (60, 90)  (sensitivity to pair correlations)
in Ref. \cite{Li11}, (54-56, 74-80), in Ref. \cite{Li10a}, (50, 52-80) in Ref. \cite{Li11a}, (38-40, 60) in Ref. \cite{Xiang12}, (36, 32-50) in Ref. \cite{Fu13}, (30, 34-38) in Ref. \cite{Song11},
and (14-18, 28) in Ref. \cite{Li11b}.

Delaroche and collaborators \cite{Delaroche10} used the 5DBH+GI, which is derived from the Gogny Interaction. 
By means of self-consistent cranking for rotation, they include the TV corrections into the 
moments of inertia. The vibrational mass coefficients are of the IB form. ZPE corrections are taken into account. There is no scaling of the energies. 
In a bench mark study, they carried out calculations for all even-even  
with $10 \leq Z \leq 100$ and $200\leq 200$. The results for the energies and $E2$ and $E0$ matrix elements for the yrast levels with $I\leq 6$, the lowest excited 
0$^+$ states, and the two next yrare 2$^+$ states accessible in the form of a  table as supplemental material to the publication. 
A thorough statistical analysis of the merits of performance has been carried out. The authors state: 
"We assess its accuracy by comparison with experiments on
all applicable nuclei where the systematic tabulations of the data are available. We find that the predicted radii
have an accuracy of 0.6\%, much better than can be achieved with a smooth phenomenological description. The
correlation energy obtained from the collective Hamiltonian gives a significant improvement to the accuracy of
the two-particle separation energies and to their differences, the two-particle gaps. Many of the properties depend
strongly on the intrinsic deformation and we find that the theory is especially reliable for strongly deformed
nuclei. The distribution of values of the collective structure indicator $R42 = E(4^+_1 )/E(2^+_1 )$ has a very sharp peak
at the value 10/3, in agreement with the existing data. On average, the predicted excitation energy and transition
strength of the first $2^+$ excitation are 12\% and 22\% higher than experiment, respectively, with variances of the
order of 40-50\%. The theory gives a good qualitative account of the range of variation of the excitation energy of
the first excited 0$^+$ state, but the predicted energies are systematically 50\% high. The calculated yrare $2^+$ states
show a clear separation between $\gamma$ and $\beta$ excitations, and the energies of the $2^+$
$\gamma$ vibrations accord well with experiment. The character of the $0^+_2$ state is interpreted as shape coexistence or 
$\beta$-vibrational excitations on the
basis of relative quadrupole transition strengths. Bands are predicted with the properties of $\beta$ vibrations for many
nuclei having $R42$ values corresponding to axial rotors, but the shape coexistence phenomenon is more prevalent."
In addition they observe that the $0^+_2$ states are generally  "too vibrational".
\begin{figure}[t]
\begin{center}
\psfig{file=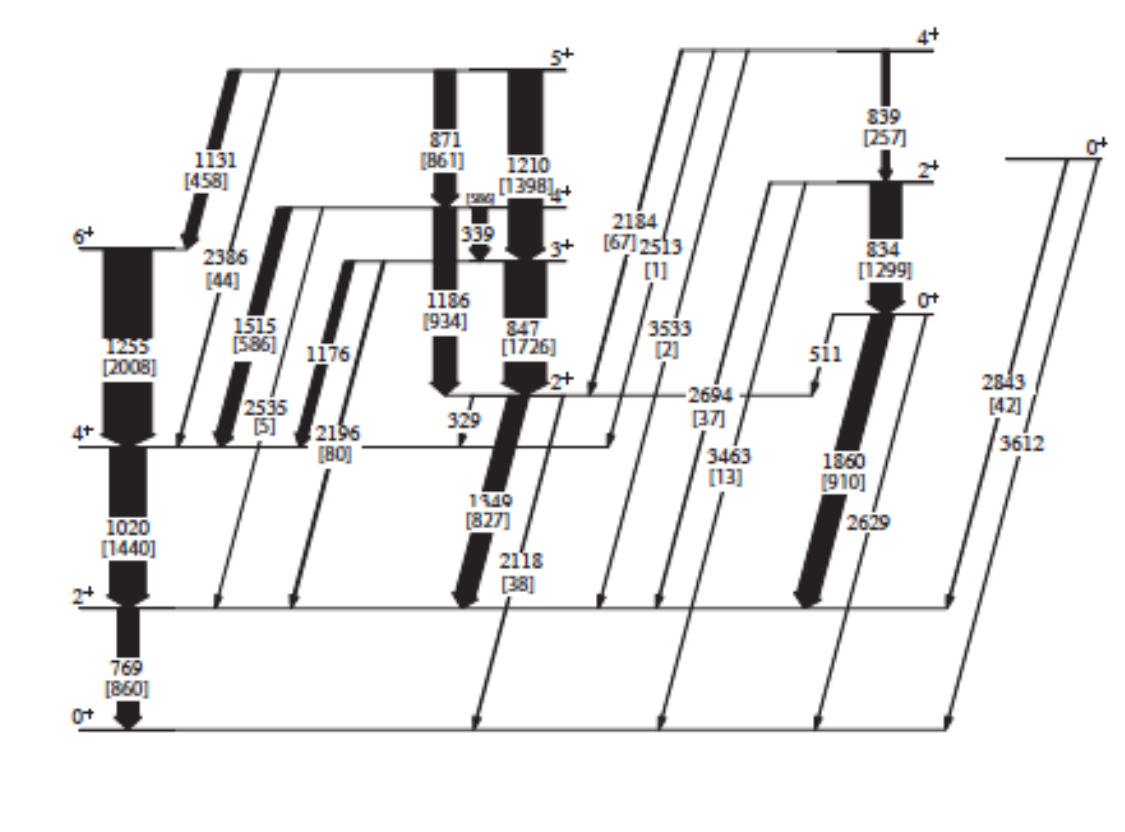,width=\linewidth}
\caption{\label{fig:102Pd_del} The spectrum of $^{102}$Pd. 
calculated by means of the 5DBH-GI \cite{Delaroche10}. 
The number in parenthesis under the transition energies (keV) are the $B(E2)$ values ($e^2$ fm$^4$) for the transitions. 
Preparation of the figure by A.D. Ayangeakaa acknowledged.  }
\end{center}
\end{figure}    

\section{Generator Coordinate Method}\label{sec:GCM}
The  Generator Coordinate Method (GCM - unfortunate coincidence with GCM for the Geometric Collective Model) constructs the
collective wave function as a superposition of the constraint mean field solutions $\vert \Psi\rangle =\int d^5\alpha f(\alpha) \vert mf, \alpha\rangle$,
where $\alpha=\{\alpha_{-2}, ...,\alpha_2\}$ is a short hand notation for the five components.
Minimizing the energy leads to the Hill-Wheeler  (HW) eigenvalue problem for the weight function $f(\alpha)$  
\begin{equation}\label{HWalpha}
 \int d^5\alpha'  \left<\alpha\vert H\vert \alpha'\right>f(\alpha')= E\int d^5\alpha' \left<\alpha\vert \vert \alpha'\right>f(\alpha'). 
  \end{equation}
The overlap kernel \mbox{$\left<\alpha\vert \vert \alpha'\right>$} in the integral equations appears because the set of mean field solutions $\vert \alpha\rangle$
represents a non-orthogonal basis. 

The Gaussian Overlap Approximation (GOA)  by Girod and Grammaticos  \cite{GC75,GC79} allows one to recast the integral equations into the form of
 the standard 5DBH differential equation.
  It approximates the overlap kernel by a Gaussian \mbox{  
$\left<\alpha_\mu\vert \vert \alpha'_\mu\right>=\exp\left[\sum_{\mu\nu}g(\bar \alpha)_{\mu\nu}\left(\alpha_\mu-\alpha_\mu'\right)\left(\alpha_\nu-\alpha_\nu'\right)\right]$},
where $\bar\alpha=(\alpha+\alpha')/2$,
and the energy kernel $\left<\alpha\vert H\vert \alpha'\right>$ by a second order Taylor expansion around  $\left<\alpha\vert \vert \alpha'\right>$. The GOA provides expressions for the ZPE. 
The mass coefficients are of the Peiers-Yoccoz (PY)  type, which are known to be smaller than the IB values. 
For this reason they are replaced by the IB expressions when the GOA is used for mapping the HW integral equations  on the 5DBH differential equations. 

The direct solution of the HW integral equations avoids the second order Taylor expansion of the energy kernel, 
which is one part of the adiabatic approximation. However using only the ground 
state configuration of the mean field solutions as HW basis assumes that the collective motion is slow enough that the coupling to excited mean field 
configurations can be neglected, which is  another part of the adiabatic approximation.
The HW equations are reformulated using the deformation parameters $\beta$ and $\gamma$ and the three Euler angles $\Omega$. The integration over the angles takes the form  
of the projection operators on good angular momentum   $P^I_{MK}$, and the integration of the deformation parameters  $a=\{\beta,\gamma\}$ is discretized.
The HW equations become
\begin{equation}\label{HWbeta}
\sum_{jK}\left<a_i\vert H P^I_{MK}\vert a_j\right>f^I_K(a_j)= E\sum_{jK}\left<a_i\vert P^I_{MK}\vert a_j\right>f^I_K(a_j). 
  \end{equation}

Restriction  to axial shapes $\gamma=0$ and $M=K=0$ considerably reduces the numerical effort.  Axial  GCM calculations have become routine for the various mean field approaches, which cannot be 
reviewed for space limitation. The references can be found in the articles about the triaxial extension to be discussed below. 
An illustrative  example is detailed study of  $^{154}$Sm in the framework of the RMF in Ref. \cite{NVR11}.  

The direct solution of the  HW is numerically very demanding because of the dimension of five. For this reason it has been applied only   to light nuclei so far. 
Bender and Heenen \cite{BH08} studied $^{24}$Mg generating the HFB mean field solutions from the SK-EDF combined with a zero-range pairing interaction.
The mean field solutions are projected on good particle number. The study focuses on the new aspects of triaxiality, like the mixing of the angular momentum components $K$
with respect to the body-fixed frame. Rodr\'iguez and Egido \cite{RE10} used  particle number projected HFB solutions generated from the Gogny interaction to carry out calculations for 
 $^{24}$Mg. Comparing with the full triaxial calculation, they observe that the ground band and the $\beta$ band on the $0^+_2$ state are rather 
 well reproduced by the axial approximation. 
 
  \begin{figure}[t]
\begin{center}
\psfig{file=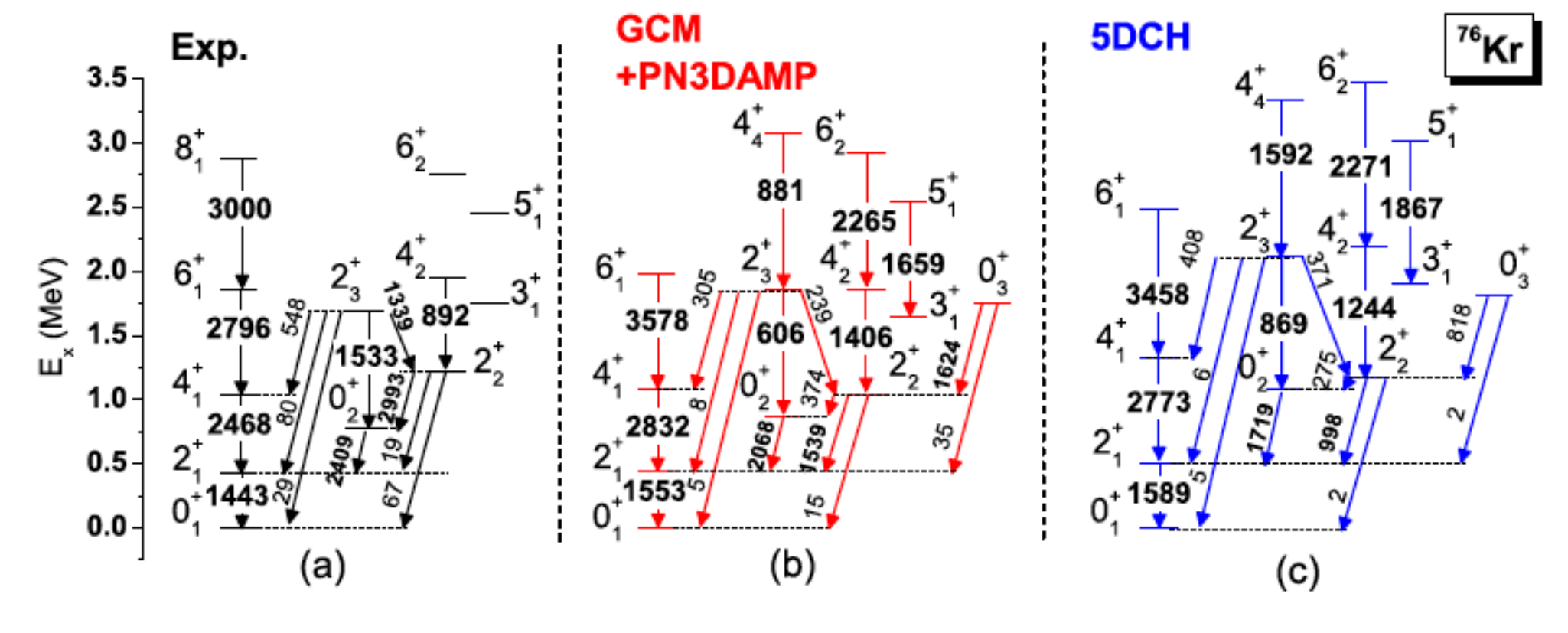,width=\linewidth}
\caption{\label{fig:76KrGCM}  Low-lying spectra and $B(E2)$ values (in $e^2~ fm^4$) of $^{76}$Kr. Results from (b) the full RMF-GCM calculation
 compared with (c) 5DCH-RMF results  and with (a) experimental data.  
Reproduced from Ref. \cite{Yao14}.}
\end{center}
\end{figure}    
 
 Yao {\it  et al. } \cite{Yao14} carried out a benchmark calculation for $^{76}$Kr, which were based on the RMF. Fig. \ref{fig:76KrGCM}
 displays the results, which are compared with   5DBH+RMF calculations using the same interaction.   
 The results agree rather well. The 5DBH spectrum is somewhat too diluted, while the GCM gives the right scale. This comes as a surprise, because earlier studies
found that the PY  moments of inertia of GCM are smaller than the IB cranking values. The reason is unclear.
  The low-lying states of $^{76}$Kr have been interpreted in terms of a prolate ground state  coexisting  with  an excited 0$^+_2$ state of smaller oblate deformation. 
 The deformation difference is manifest by the spacing of the two rotational bands built on the band heads. Both approaches reproduce the experiment in considerable detail, 
 which is also the case for the 5DBH+SK (Fig. 10 of Ref.\cite{Fu13}) and 5DBH+GI (Fig. 16 of Ref. \cite{Clement07})  calculations for this nucleus. Tab. II of Ref. \cite{Yao14} compares calculated spectroscopic quadrupole moments
 with experiment. The figures and the table represent the state of art of microscopic approaches based
  on self-consistent mean field approaches.    
 
 \begin{figure}[t]
\begin{center}
\psfig{file=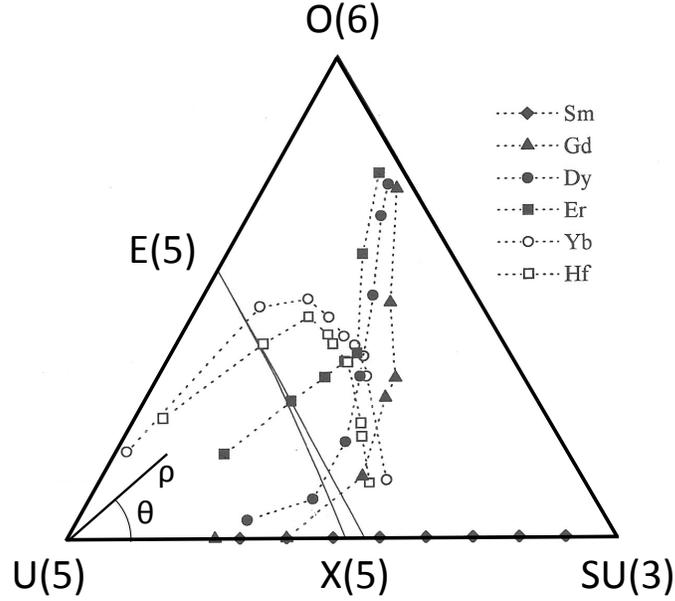,width=\linewidth}
\caption{\label{fig:Triangle} The IBM parameters of selected isotope chains arranged in the Symmetry Triangle. The symmetry limits
of the subgroups are indicated at the three corners. The  "transition point symmetry" X(5) marks 
the transition from the vibrational regime (limit U(5))  to axial deformation (limit SU(3)  
and E(5) marks the transition to the $\gamma$ - soft deformed regime (limit O(6)).   Adapted from Ref. \cite{Zamfir03}.}
\end{center}
\end{figure}    

 \section{The Interacting Boson Model }\label{sec:IBM}
The collective quadrupole mode is described by means of 
the creation operators for nucleon pairs with spins 0 and 2, which are called s- and d- bosons and denoted by $s^{\dagger}$ and $d^{\dagger}$,
respectively \cite{IA87}. 
They form the closed Lie algebra of the $\grpsu{6}$ group. 
In this review only the simplest version,  the "Extended Consistent Q Formulation" of  IBM-1 of Warner and Casten   \cite{WC83}, is presented.
The Hamiltonian  contains  two IBM parameters and the energy scale, such that:
\begin{equation}
\label{eqn:HIBM}
H_{IBMA}(\zeta , \chi)= g\Big((1-\zeta) \hat{n}_d-\frac{\zeta}{4N_B} \hat{Q}^{\chi}\cdot\hat{Q}^{\chi}\Big),
\end{equation}
where $\hat{n}_d=d^{\dagger}\cdot\tilde{d}$, and $\hat{Q}^{\chi}_\mu=[s^{\dagger}\tilde{d}+d^{\dagger}s]^{(2)}_\mu+\chi[d^{\dagger}\tilde{d}]^{(2)}_\mu$.
The Hamiltonian is diagonalized within the space of  fixed number of bosons, $N_B=n_s+n_d$, which is taken to be
 half the number of valence nucleons. 
 The matrix elements of the charge quadrupole moments are taken to be proportional to $\hat{Q}^{\chi}$, with an effective boson 
 charge fixing the scale. As in the case of the GCM, the two parameters $\zeta$ and $\chi$ determine the character of the collective states. 

It has become custom to map the structure of the collective states on the "symmetry triangle", which is a polar plot of  $\zeta$ and $\chi$
with the radius $\rho$ and the angle $\theta$ defined by
\begin{eqnarray}
\rho=\frac{\zeta\sqrt{3}}{\sqrt{3}\cos \theta_\chi-\sin \theta_\chi},~~~
\theta=\frac{\pi}{3}+\theta_\chi, ~~~\theta_\chi=\frac{2\pi}{3\sqrt{7}}\chi.
\end{eqnarray}
The three corners of the triangle are parameter combinations that generate additional symmetries with respect to the subgroups U(5), O(6), SU(3), 
 which correspond to  the harmonic vibrator, the $\gamma$ -independent, axial rotor limits of the GCM approach. It is an attractive feature of the IBM that
 simple algebraic expressions describe the energies and reduced transition probabilities of the three symmetry limits.
The two-parameter triangle is used to classify nuclei, where the concept of "Quantum Phase Transitions" is invoked.  
Order parameters $\beta_B,\gamma_B$ are defined by mapping  
the IBM Hamiltonian on a basis of coherent states which represent a condensate of  d-bosons. 
\begin{equation}
\label{Coherent}
\big\vert N_B,\beta_B,\gamma_B \big>=\frac{1}{\sqrt{N!}} \hat{B}^{+N_B}\big\vert0\big>, 
\end{equation}
where
\begin{equation}
\hat{B}^{+}=s^++\beta_B\Big(cos(\gamma_B)d^+_0+\frac{sin(\gamma_B)}{\sqrt{2}}(d^+_2+d^+_{-2})\Big).
\end{equation}
The expectation value of the IBM Hamiltonian with the coherent state (\ref{Coherent}) has been given by Ginoccio and Kirson \cite{GK80},
\begin{displaymath}\label{EIBM}
 E_{IBM}(\beta_{B},\gamma_B)=\big< N_B,\beta_B,\gamma_B \big\vert H_{IBM} \big\vert N_B,\beta_B,\gamma_B \big> 
\end{displaymath}
\begin{displaymath}
=c_E \bigg(\frac{\frac{-5}{4}\zeta+\big((1-\zeta)N_B-\frac{1}{4}\zeta(1+\chi^2)\big)\big(\beta_{B}\big)^2}{1+(\beta_{B})^2}
%\end{displaymath}
%\begin{displaymath}
-\Big(\frac{\zeta(N_B-1)(\beta_{B})^2}{(1+(\beta_{B})^2)^2}\Big) 
\end{displaymath}
\begin{equation}
\times \Big(1-\sqrt{\frac{2}{7}}\chi \beta_{B} cos(3\gamma_B)+\frac{\chi^2}{14}(\beta_{B})^2\Big)\bigg).
\end{equation}
Border lines between the regimes characterized by the symmetry limits are defined by applying the Ehrenfest classification of instabilities of
thermodynamic potentials  to  the energy
function $E_{IBM}(\beta_{B},\gamma_B)$, which are shown  in Fig. \ref{fig:GCMmap}.
Only the thermodynamic limit of an infinite system leads to a sharp phase boundary.
The instabilities of  $E_{IBM}(\beta_{B},\gamma_B)$ demarcate the center of a cross-over region between the three regimes.
  In order to describe the instabilities by simple algebraic expression, Iachello \cite{E5X5} invoked the discussed solutions 
of the BH with a schematic square well potential  (c. .f Sec. \ref{sec:GCMphen}).   These are commonly referred to as the 
X(5) and E(5) "transition point symmetries" although they are  no algebraic symmetries that correspond to a symmetry group like 
 the corners of the triangle. 

\begin{figure}[t]
\begin{center}
\psfig{file=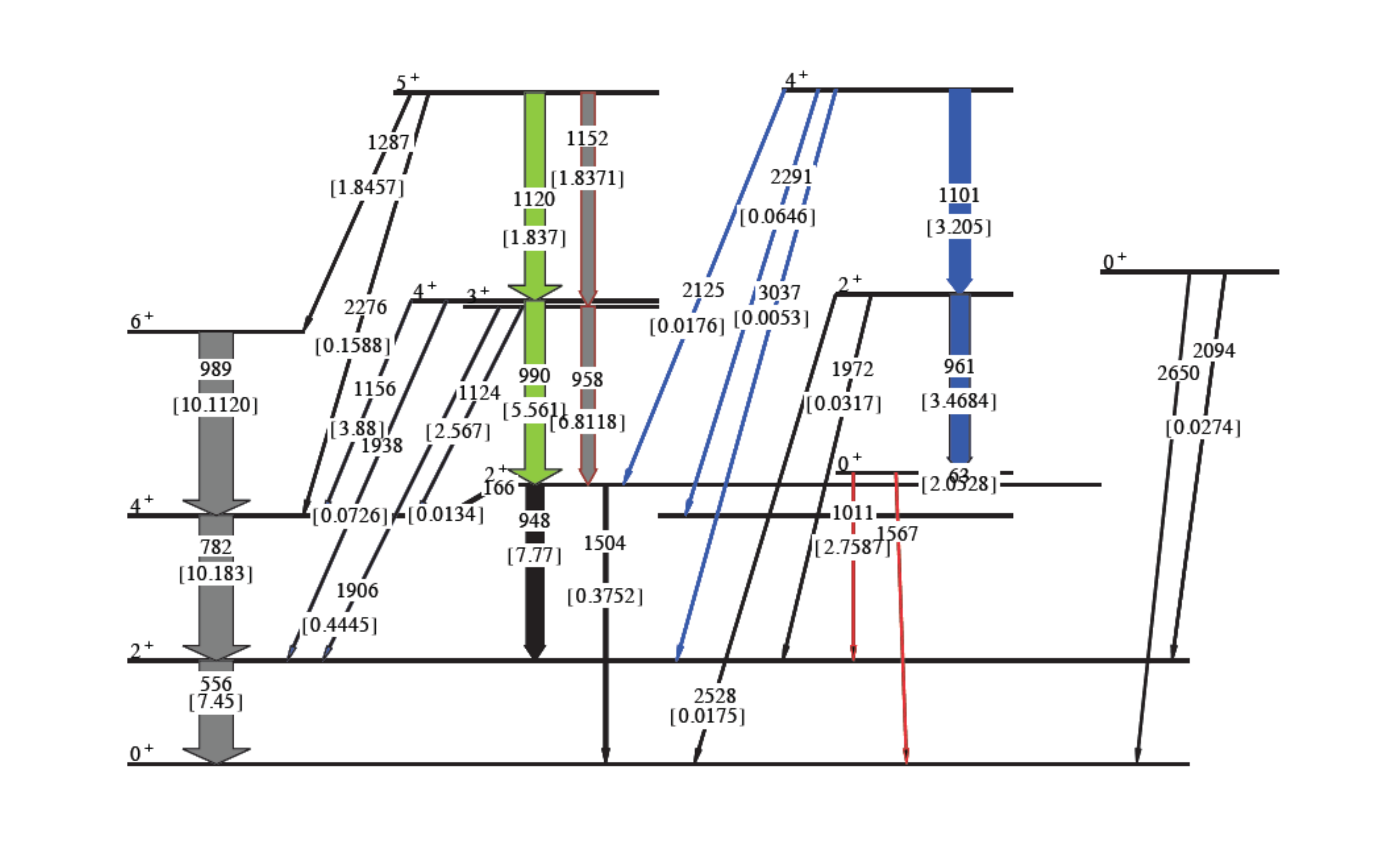,width=\linewidth}
\caption{\label{fig:102PdIBM} IBM-1 fit to the spectrum of $^{102}$Pd.   The number in parenthesis under the transition energies (keV) are the $B(E2)$ values ($e^2$ fm$^4$) for the transitions.
The IBM parameters are $\zeta=0.61,~\chi=0.58, N_B=5$. Preparation of the figure by W. Li is acknowledged. } 
\end{center}
\end{figure}

The two-parameter fits usually well account for the relative energies of the   $2^+_1$, $4^+_1$, $6^+_1$, $2^+_2$, $3^+_1$, $4^+_2$, $0^+_2$, $4^+_3$ states
and the relative $B(E2)$ for the transitions between them. The quality of the fits is comparable with the two-parameter version of the GCM discussed in Section \ref{sec:GCMphen}.
Fig. \ref{fig:102PdIBM} shows an IBM-1 fit as an example, which is to be compared with the experiment in Fig. \ref{fig:102Pd_del}. Like the GCM, the standard IBM
encompasses  only the limit of $\gamma$-instability. To stabilize triaxility, the IBM Hamiltonian has been complemented by a third order term in   $\hat{Q}^{\chi}_\mu$ \cite{Isacker81}.
The extension is well exposed in Ref. \cite{Stefanescu07}, which discusses the staggering of the $\gamma$ band in $^{110,112,114}$Ru as a signature for  triaxiality.
Since the appearance of triaxiality in the IBM framework is analog to the one in the framework of the BH discussed in Sect. \ref{sec:triaxiality}, it will not be further reviewed.  

\subsection{Boson Number Limit}\label{sec:BNcut}

\begin{figure}[t]
\begin{center}
\psfig{file=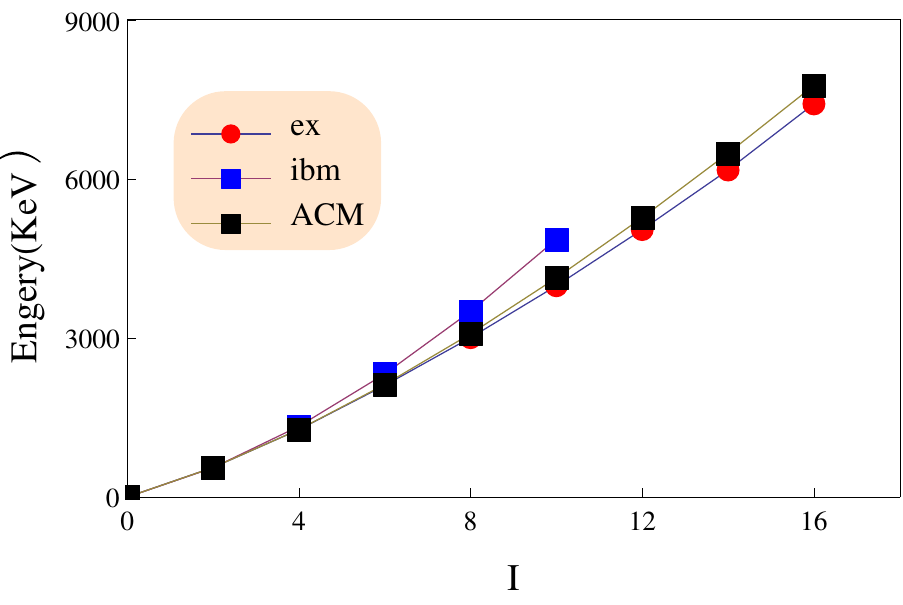,width=6cm}
\psfig{file=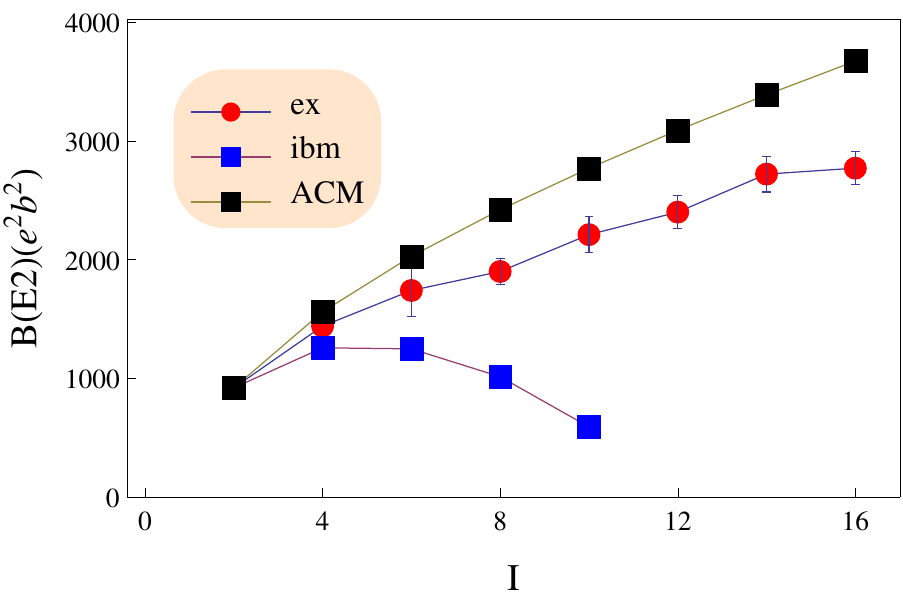,width=6cm}
\caption{\label{fig:102PdEBE2} Energies (left) and $B(E2, I\rightarrow I-2)$ values (right) of $^{102}$Pd.  The GCM calculation, denoted by ACM, is the same as the one shown in
Fig. \ref{fig:102PdACM} and the IBM calculation is the same as in Fig. \ref{fig:102PdIBM}. Data from Ref. \cite{102PdTidalPRL}.} 
\end{center}
\end{figure}

The IBM has been conceived as an approximation to the Shell Model. The configuration
space of the valence nucleons between two closed shells is truncated  to the subspace of pairs coupled to spin
zero and two, which are then mapped to the space of the $s$ and $d$ bosons. The number $N_B$  of such pairs is  taken
one half of the number of valence particles below the middle of the open shell and one half of the valence holes above the middle.
The finite boson number is considered to be the major difference between  the IBM and the phenomenological versions of the BH, which can also
be cast into algebraic form (see the profound discussion in the textbook by Rowe and Wood \cite{RW}).   The consequences of the boson number
limit have not been well studied because the IBM is usually  applied to  low-spin states, which correspond to bosons numbers far below the limit.  
The recent measurement of the life times of the yrast states in $^{102}$Pd in Ref. \cite{102PdTidalPRL} provides such test. 
As illustrated by Fig. \ref{fig:102PdEBE2},
the yrast levels form a regular sequence of collective excitations with increasing $B(E2, I\rightarrow I-2)$ values up to $I=14$. 
The data are compared with the GCM and IBM calculations that give the spectra shown in Figs. \ref{fig:102PdACM} and  \ref{fig:102PdIBM}. According to the IBM counting rule,
$^{102}_{46}$Pd$_{56}$ has a boson number of $N_B=5$, four valence  proton holes and six valence neutrons with respect to $Z=N=50$. The finite number of bosons 
limits the regular yrast sequence at $I=10$ , which is the maximum that can be generated by five  $d$ bosons. The $B(E2)$ values decrease toward
the limit where they are zero. These consequences of the finite boson number are in clear contradiction with experiment. The GCM calculation, which does not assume a boson limit
reproduces the experiment quite well. 
 Figs. \ref{fig:102PdACM} and  \ref{fig:102PdIBM} demonstrate that both the GCM and the IBM reproduce the low-spin part
of the spectrum, where the boson cut-off is of minor importance,  with comparable accuracy. The example shows that the collective angular momentum
cannot originate  from the spherical valence particles and holes coupled to spin zero and two only. 
Testing the consequences of the finite boson number assumption of the IBM for other cases would be interesting.

\subsection{Mean field mapping}
IBM assumes that the $s$ and $d$ bosons 
 represent valence nucleon pairs in spherical orbitals that are coupled to spin 0 or 2.  A microscopic derivation  of the IBM parameters starting from this  concept has not
 been succeeded  for nuclei located far in the open shell. 
An alternative approach by Nomura and collaborators \cite{Nomura08} and Bentley and Frauendorf  \cite{Bentley11} has provided 
encouraging results. The IBM parameters are determined by adjusting the IBM potential energy surface (PES) $E_{IBM}(\beta_{B},\gamma_B)$ given by Eq. (\ref{EIBM}) 
to the PES $E_{mf}(\beta_F,\gamma_F)$ calculated by means of constraint mean field theory.  For simplicity,   the version of Ref.   \cite{Bentley11} is presented,
which  determines the parameters of the IBM-1 Hamiltonian (\ref{eqn:HIBM}). The mean field  PES  is calculated by means of the micro-macro approach.
 The mapping contains four free parameters, the energy scale $c_E$, the deformation scale $\beta_B=c_B\beta_{F}$, and the IBM parameters   $\chi$ and $\zeta$.  
The deformation scale $c_B$ is fixed by the requiring that the location of the  minimum of $E_{IBM}(\beta_{B},\gamma_B)$ agrees with the location of the minimum
of $E_{mf}(\beta_F,\gamma_F)$. The energy scale $c_E$ is fixed by the energy of the 2$^+_1$ state, which is calculated by means of the cranking model (see Section \ref{sec:tidal})
or taken as  the experimental value. The IBM parameters $\chi$ and $\zeta$ are found   by fitting  $E_{IBM}(\beta_{B},\gamma_B)$ to   $E_{mf}(\beta_F,\gamma_F)$
in the region $E<$ 1MeV, which is essential for the low-energy collective states, and assuming $\gamma_B=\gamma_F$. 
Ref. \cite{Bentley11} calculated the IBM  parameters and spectral properties for the nuclides with  $(Z,~N)=$
 (36, 40-50), (42, 56-66), (46, 56-70), (48, 60-70), (64, 88-98), (66, 90-100), (68, 88-100). The method describes the spectral structure of the nuclei 
amenable to the IBM parametrization fairly well. If $c_E$ is fixed by the experimental energy of the 2$^+_1$ state the spectrum tends to be too stretched. Using the cranking
value results gives a scale close to experiment.    

Nomura and collaborators used the IBM-2 version. They introduced an additional term $\propto I(I+1)$ the coefficient of which is determined by a cranking calculation. 
In fitting $E_{IBM}(\beta_{B},\gamma_B)$, they   assumed that the neutron and proton parameters are equal. Calculations mapping $E_{mf}(\beta_F,\gamma_F)$
generated from two versions of a constrained Skyrme EDF   have been carried out for the nuclides with  $(Z,~N)=$  
 (62, 84-94), (56, 54-66), (54,  54-66), (44, 54-80), (46, 54-80), (74-76, 130-140) in Ref. \cite{Nomura10}, (62, 84-96), (92, 144-146) in Ref. \cite{Nomura11}, (64-66, 84-96) in Ref. \cite{Kotila12}. 
 Calculations using the constraint HFB applied to the 
Gogny interaction have been carried out for the nuclides with  $(Z,~N)=$  
(78, 102-120) in Ref. \cite{Nomura11Ga}, (74-76, 114-120) in Ref. \cite{Nomura11Gb}, (70-78, 110-122) in Ref. \cite{Nomura11Gc}.
The experimental energy scale and the spectral structure is well reproduced for the nuclei amenable to the IBM phenomenology. Like in Ref. \cite{Bentley11}, the
0$^+_2$ states are frequently predicted too high. This concerns not only the very low-lying 0$^+_2$ states that are excluded as "intruder states" from the 
practiced IBM phenomenology, but also many of the    0$^+_2$ states in well deformed nuclei.   The 0$^+_2$ states in  X(5) nuclei are reproduced best, because they have 
the character of a soft anharmonic vibration in the $\beta$ degree of freedom.   
In Ref. \cite{Nomura11RMF}, Nomura {\it et al.} compared for $^{192-196}$Pt the results obtained by mapping $E_{mf}(\beta_F,\gamma_F)$ from RMF 
to IBM-2 with 5DBH-RMF calculations based on the same RMF. The IBM-2 generates a quasi $\gamma$ band with the pronounced even$I$-low staggering of $\gamma$-soft
nuclei whereas      5DBH-RMF, consistent with experiment, gives very small staggering signaling the stabilization of the triaxial shape (cf. Sec. \ref{sec:triaxiality}).
One may expect that calculating the parameters of an IBM Hamiltonian augmented by a third order term  by means of the mapping procedure will fix this deficiency. 

The success of the mapping technique appears puzzling from the point of view of the ATDF approach, because the values of the mass parameters are determined by the 
the potential energy surface. The relation, which has not been made explicit yet, originates from the algebraic structure of the IBM.  However, Fig. \ref{fig:VBbbWf124Xe} 
does not reveal an obvious correlation between $B_{\beta\beta}$ and $V$, which is an example for the general situation.   

\section{Non-adiabatic approaches}\label{sec:noadia}

The various approaches in Sections \ref{sec:ATDMF} - \ref{sec:IBM} presume that the quadrupole mode is decoupled from the quasiparticle excitations.  
Fig. \ref{fig:decoherence} schematically illustrates that this presumption does not hold far. The 0$^+_2$ member of the two-phonon triplet is usually close to the lowest 
0$^+$ two-quasiparticle excitation and some coupling is expected,. (The situation for the $\beta$ vibration in  well-deformed nuclei is analogous. )
The vicinity of the quasiparticle excitations is the reason why  the collective models perform poorest for the 
0$^+$ excitations. The low-$I$ members of the higher phonon
multiplets become progressively mixed with the quasiparticle excitations, as seen in the E2 strength distribution of the Shell Model calculation on the right 
side of Fig. \ref{fig:decoherence}.  Very soon soon they are 
"dissolved  in the  sea of  quasiparticle excitation", which means they cannot be 
described by a coherent collective wave function.  According to BH phenomenology, the 0$^+_2$ states  of well deformed nuclei are expected to
be a $\beta$ vibration around the axial equilibrium shape. However, the careful analysis  by Garrett \cite{Garrett01}  demonstrated that the properties of the 0$^+_2$ excitations deviate
qualitatively from the characteristics of a  $\beta$ vibration. Only the  the very collective   0$^+_2$ states in transition point nuclei come close to the predictions of the 
phenomenology  or, alternatively, can be interpreted in terms of  shape coexistence. 

For  the yrast states of the multiplets
the sea of quasiparticle excitations is approached more slowly, because the level density increases as a function of the distance from the yrast line.  In well-deformed nuclei, the low part of
the yrast line is just the rotational band built on the deformed ground state. Around $I=10$ this ground  (g) band encounters  the s band, which is built on a non-collective excitation  composed
of  two high-j quasiparticles which align their spin. This interplay between the collective rotational degree of freedom and the quasiparticle degrees of freedom can described in considerably detail 
in the framework of the rotating mean field approaches. The deformation of the nucleus is considered as static. Depending on the specific version it is 
optimized for each angular momentum ("Total Routhian Surfaces" TRS, "Cranked Nilsson Strutinsky" approach CNS)  or kept constant ("Cranked Shell Model" CSM). 
The coupling between the collective rotation and the quasiparticle motion is taken into account in a non-adiabatic way.
The present article will not address this field (see  Frauendorf \cite{RMP} and Satu\l a and Wyss \cite{WS05} for recent reviews).  Two approaches for near-yrast states will be reviewed:
the tidal wave concept that allows one to describe  
non-adiabatic coupling between the collective quadrupole mode and the quasiparticle modes  in the vibrational and transitional regime and the Triaxial Projected Shell Model that
allows one to take into account the collective $\gamma$ mode.                 

\subsection{Tidal Waves}\label{sec:tidal}
 Frauendorf, Gu and Sun  introduced the Tidal Wave concept in Ref. \cite{FGS10} and Ref. \cite{FGS11}, which contains  complimentary material. 
 \begin{figure}[t]
\begin{center}
\psfig{file=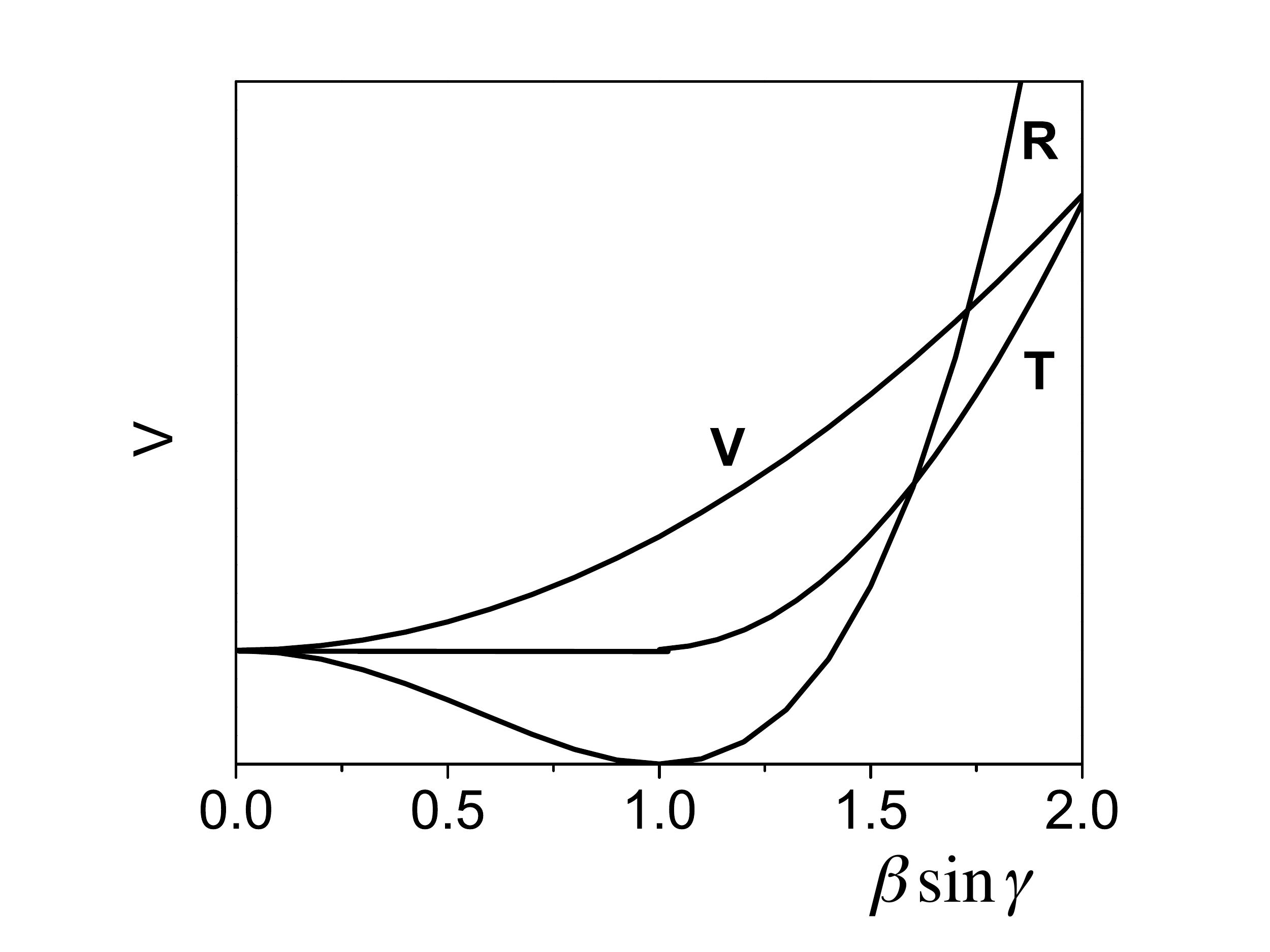,width=6cm}
\psfig{file=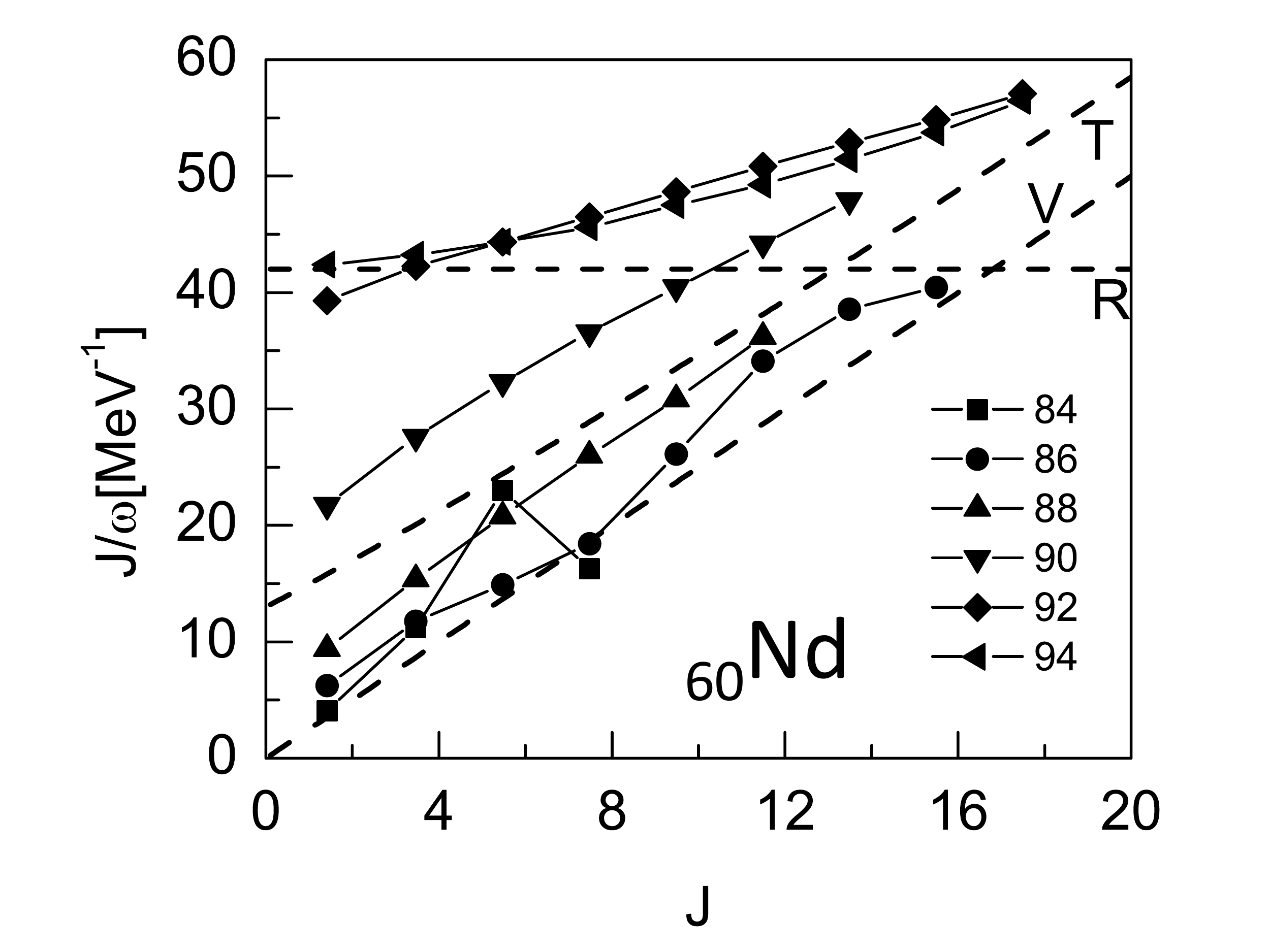,width=6cm}
\caption{\label{fig:NdJIpots} Left panel: The potential $V(\beta_e, \gamma_e)$
along the path of the deformation minimum.  Right panel:
Experimental moments of inertia of the ground bands. 
of the Nd isotopes, where \mbox{$\omega=(E(I)-E(I-2))/2$} and $J=I$.  From Ref. \cite{FGS10}.   } 
\end{center}
\end{figure}

 \subsubsection{Phenomenology}
 
 Consider the classical BH    (\ref{BHclass}) with phenomenological mass parameters  $B_{\beta\beta}=B_{\gamma\gamma}=B_i=\sqrt{5}/2D$.  
 Uniform rotation about the medium (m-)  axis, which has 
 the maximal moment of inertia, has the lowest energy for a given angular momentum, i. e. it corresponds to the yrast state when quantized. 
The choice of the interval $2\pi/3 \leq \gamma \leq \pi/3$ makes the m-axis the 3-axis  of quantization.
The location of the surface in spherical 
coordinates is given by
\begin{equation}  \label{tidalclass}
R(t)=R_o[1+\sqrt{2}\beta\sin\gamma\cos(2\phi-2\omega t)Y_{22}(\vartheta,\phi=0).
\end{equation}  
The deformation parameters $\beta$ and $\gamma$ do not depend on time, 
 because any time time dependence involves additional kinetic energy. 
 As discussed in detail in Ref. \cite{FGS10}, their values are given by minimizing the energy 
 \begin{equation}\label{Ebeta}
E(\beta,\gamma)=\frac{J^2}{2{\cal J}(\beta,\gamma)}+V(\beta,\gamma),~~
{\cal J}=4B\beta^2\sin^2\gamma.
\end{equation}
\begin{figure}[t]
\begin{center}
\psfig{file=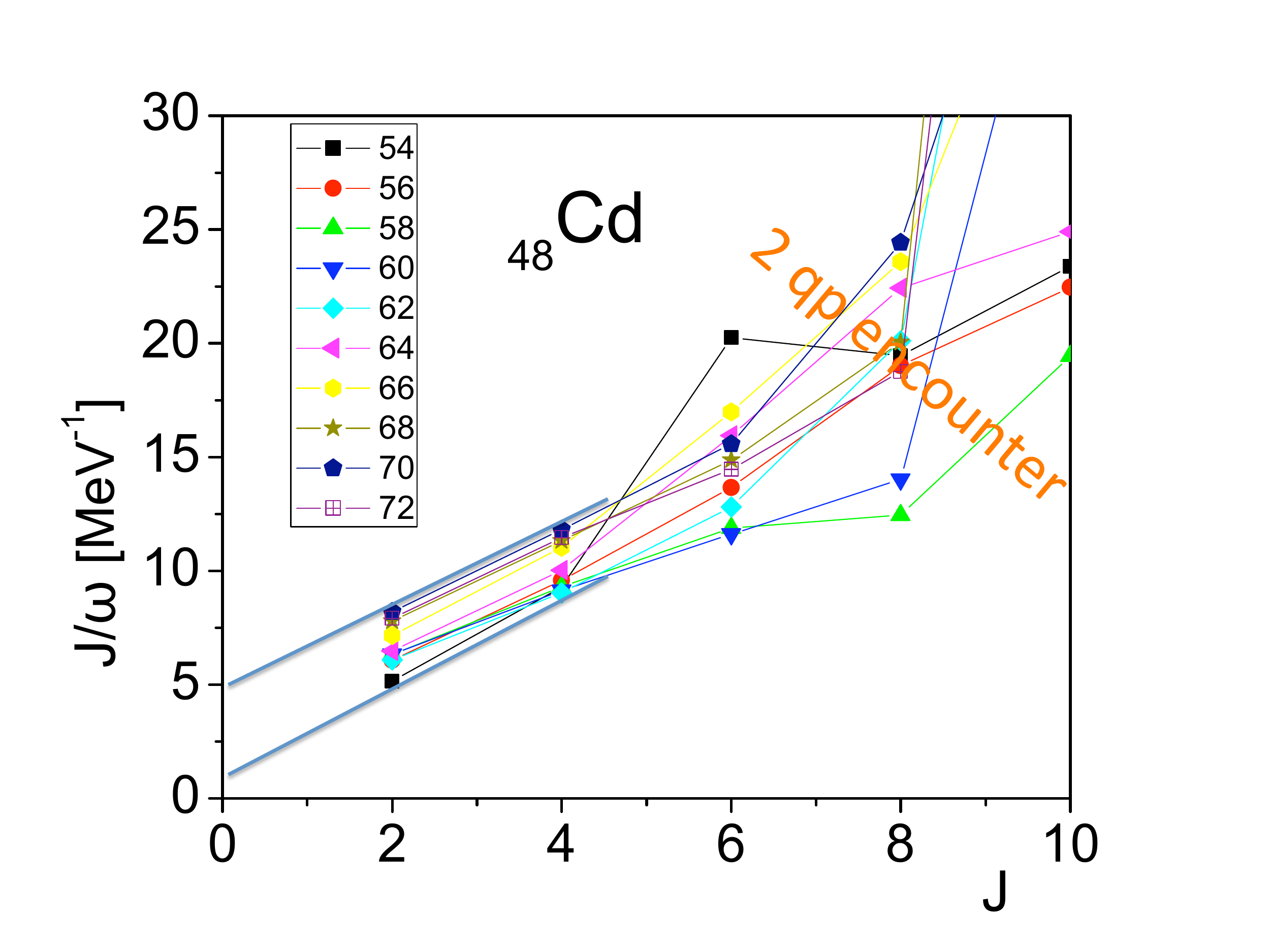,width=6.2cm}
\psfig{file=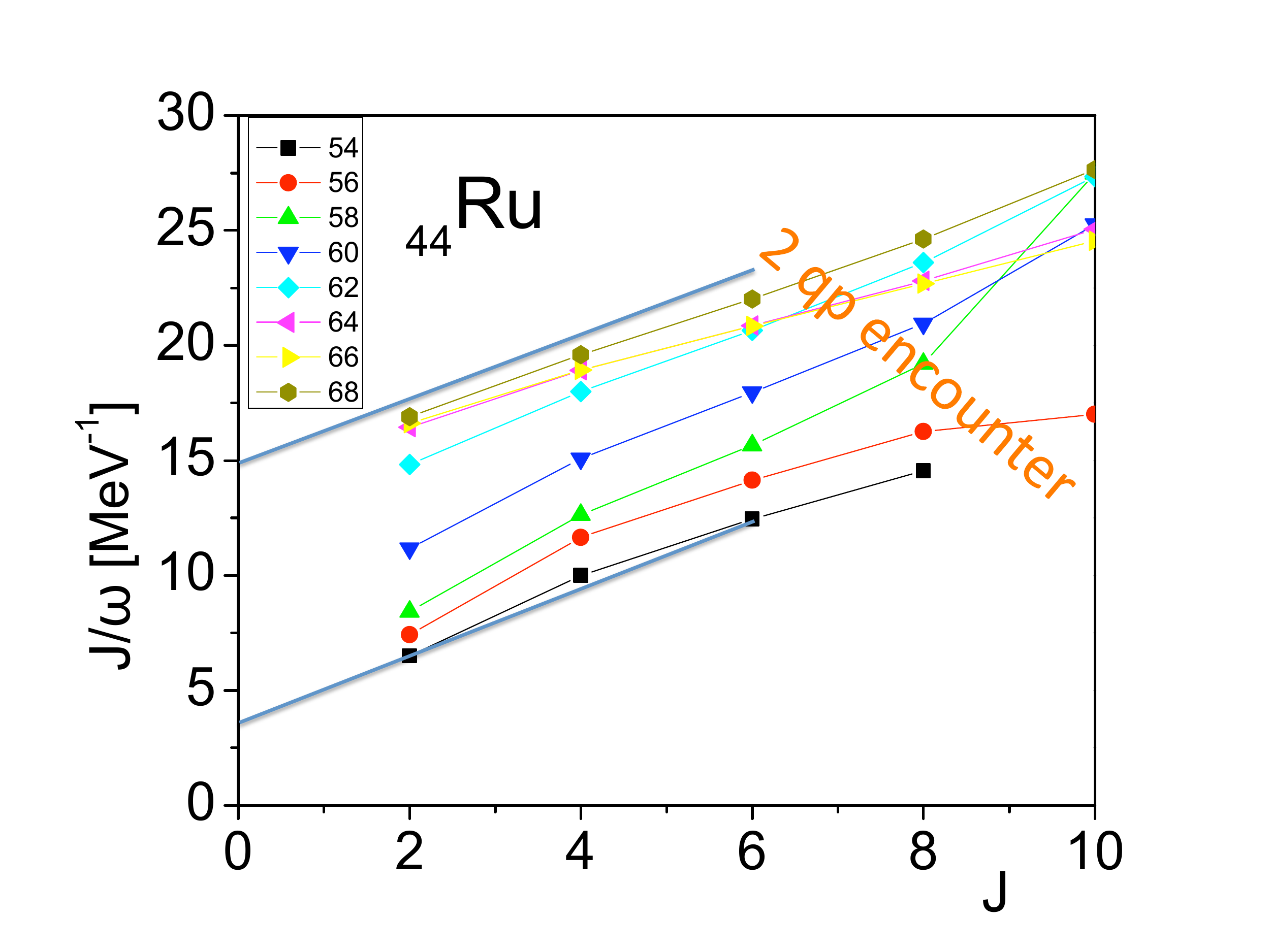,width=6.2cm}\\
\psfig{file=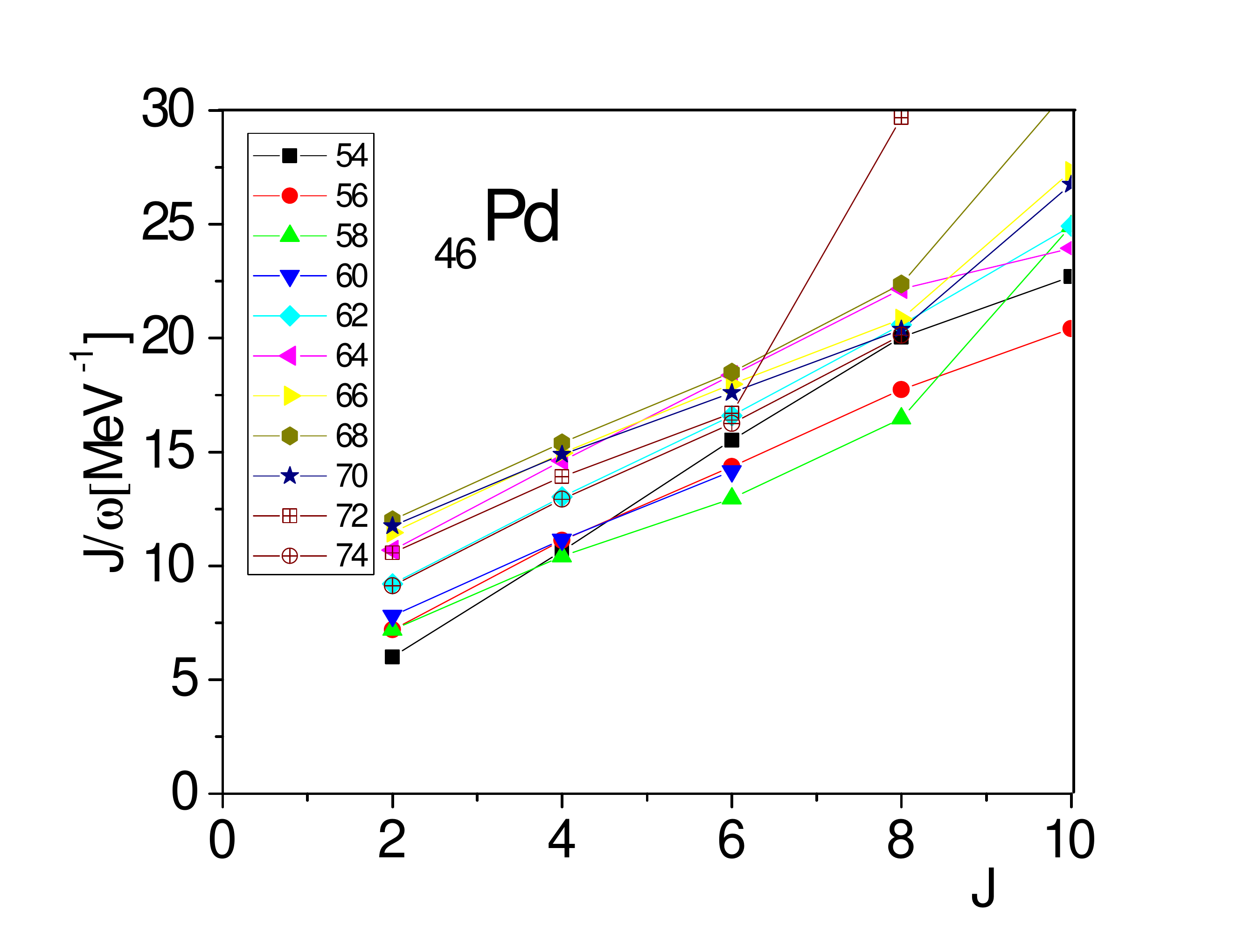,width=6.2cm}
\caption{\label{fig:CdRuJIsys} 
Experimental moments of inertia of the ground bands. 
of the Cd, Pd, and Ru isotopes, where \mbox{$\omega=(E(I)-E(I-2))/2$} and $J=I$.    } 
\end{center}
\end{figure}    
Let us discuss some cases in more detail. \\
-Harmonic vibrator:(V in  Fig. \ref{fig:NdJIpots}) $V=\frac{C}{2}\beta^2$ \\
Minimizing the energy one finds
\begin{eqnarray}
\gamma_e=\frac{\pi}{2},~ \beta^2_e=\frac{J}{2\sqrt{BC}},~ {\cal J}=4B\beta^2_e=\frac{2J}{\sqrt{BC}},\\
\omega=\frac{J}{{\cal J}}=\frac{1}{2}\sqrt{\frac{C}{B}},~E=\omega J=\Omega\frac{J}{2}=C\beta^2_e.
\end{eqnarray}
The wave travels with an angular velocity $\omega$ being one half of the oscillator frequency
$\Omega$. The angular momentum is generated by only  increasing the deformation $\beta^2$, 
which increases the moment of inertia, which is a linear function of $J$. 
This mode has been called  "tidal wave", because it has wave character: the energy and angular momentum increase with the wave amplitude while
the frequency stays constant.\\
-Axial rotor (R in Fig. \ref{fig:NdJIpots})): The potential has a minimum at $\beta_0$, where \mbox{$V\approx C(\beta-\beta_0)^2/2$}
and is stiff around $\gamma_e=\pi/3$.\\
 The minimization gives
\begin{equation}
 \beta_e=\beta_0(1+\frac{2J^2}{{\cal J}_0C\beta_0^2}+O(J^4)),~~
{\cal J}={\cal J}_0(1+\frac{8J^2}{{\cal J}_0C\beta_0^2}+O(J^4)).
\end{equation}
The moment of inertia is a slowly increasing quadratic function of $J$. Energy and
angular momentum increase  (mainly) due to growth of the angular velocity $\omega$.\\
-Transitional nucleus (T in Fig. \ref{fig:NdJIpots})):
The yrast energies are well approximated  ($J=I$) by the expression
\begin{equation}\label{Ebr}
{\cal J}=\Theta_0+\Theta_1J,~~\omega=\frac{J}{{\cal J}}, ~~E(J)=\int_0^J{\omega(J)dJ}.
\end{equation}
Assuming that the energies result from minimization of the energy  (\ref{Ebeta}), the potential $V(\beta_e,\gamma_e)$  
is given in parametric form by
 \begin{equation}\label{Vbr}
\beta_e(J)\sin\gamma_e(J)=\sqrt{\frac{\Theta_0+\Theta_1 J}{4B}},~~
V(J)=\int_0^J{\omega(J)dJ}-\omega(J)J/2,
\end{equation}   
 which has an intermediate form T in Fig. \ref{fig:NdJIpots}.   As expected for a transitional nucleus,
the angular momentum is gained  by increasing both 
$\cal{J}$ and $\omega$. From a vibrational
 perspective, the increase of $\omega$ reflects the anharmonicity
of the motion, from the rotational perspective, the increase
of ${\cal J}$ reflects the softness of the rotor. 

The chain of Nd - isotopes in Fig. \ref{fig:NdJIpots} displays the transition 
from a vibrational (V - small $\Theta_0$) to a rotational (R - large $\Theta_0$) yrast sequence. 
The value of $\gamma_e$ cannot be inferred from the yrast sequence only. In the case of the Nd isotopes 
the high energy of the $\gamma$ band points to a near-axial value close to $2\pi/3$ (X(5) type). 
For $N=84$, the vibrational sequence is short and somewhat irregular, which indicates that
the underpinning fermonic structure seems through the  weakly collective  mode. For $N=86$, the increased
collectivity smoothes out most of the irregularity, which disappears for $N=88$.
The Cd chain in Fig. \ref{fig:CdRuJIsys} has small values of $\Theta_0$ as expected for nuclei close to
the vibrational limit. The quasiparticle excitations are early encountered. The isotonic Ru chain has
transitional character with 5 MeV$<\Theta_1<$15 MeV. The low $E(2^+_2)$ energy and the pronounced even-$I$-up
staggering of the $\gamma$ band point to $\gamma_e\approx \pi/2$ (E(5) type).   It is obvious that the
encounter of the quasiparticle excitations does not correlate with the pair counting rule of IBM.

\begin{figure}[t]
\begin{center}
\psfig{file=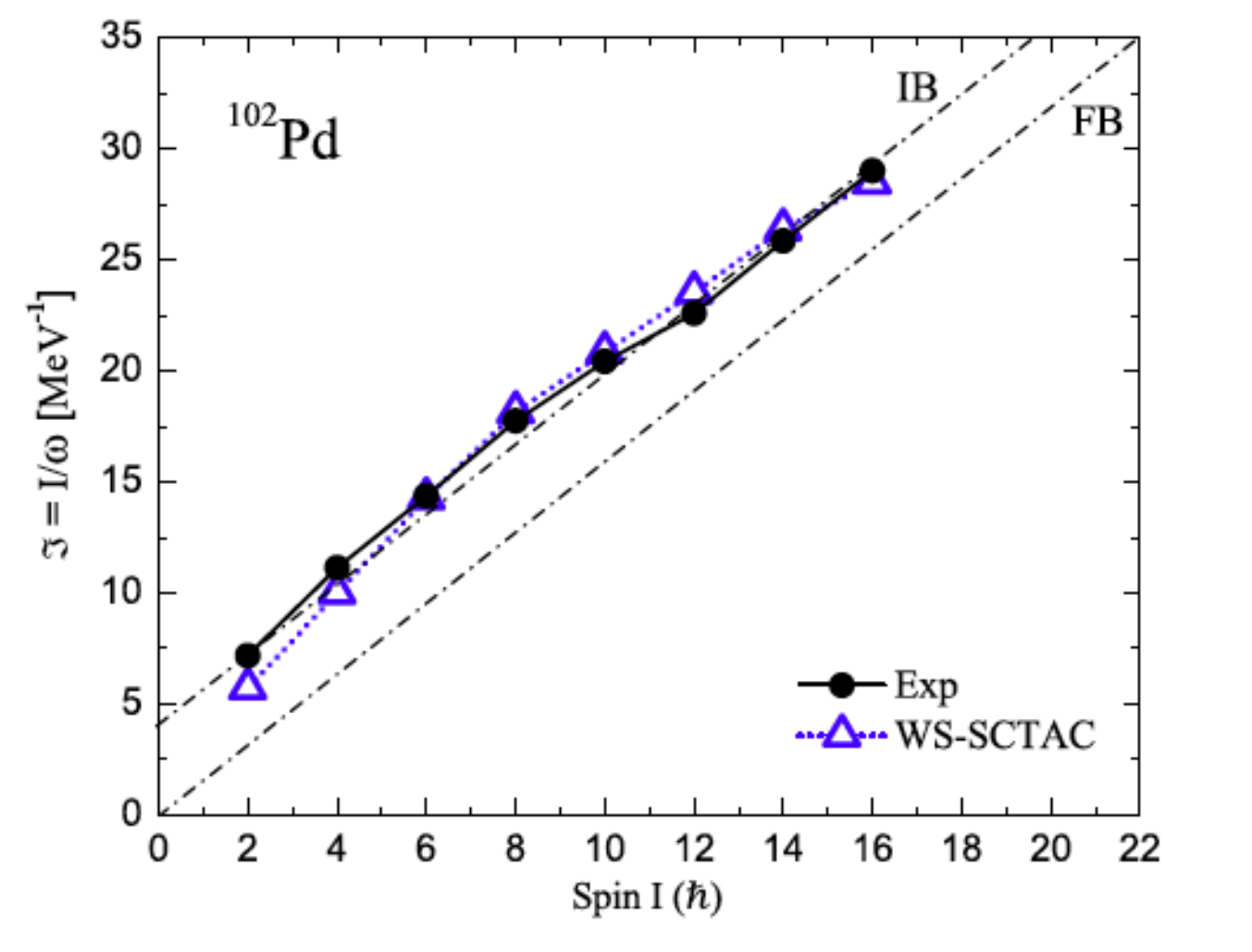,width=6.3cm}
\psfig{file=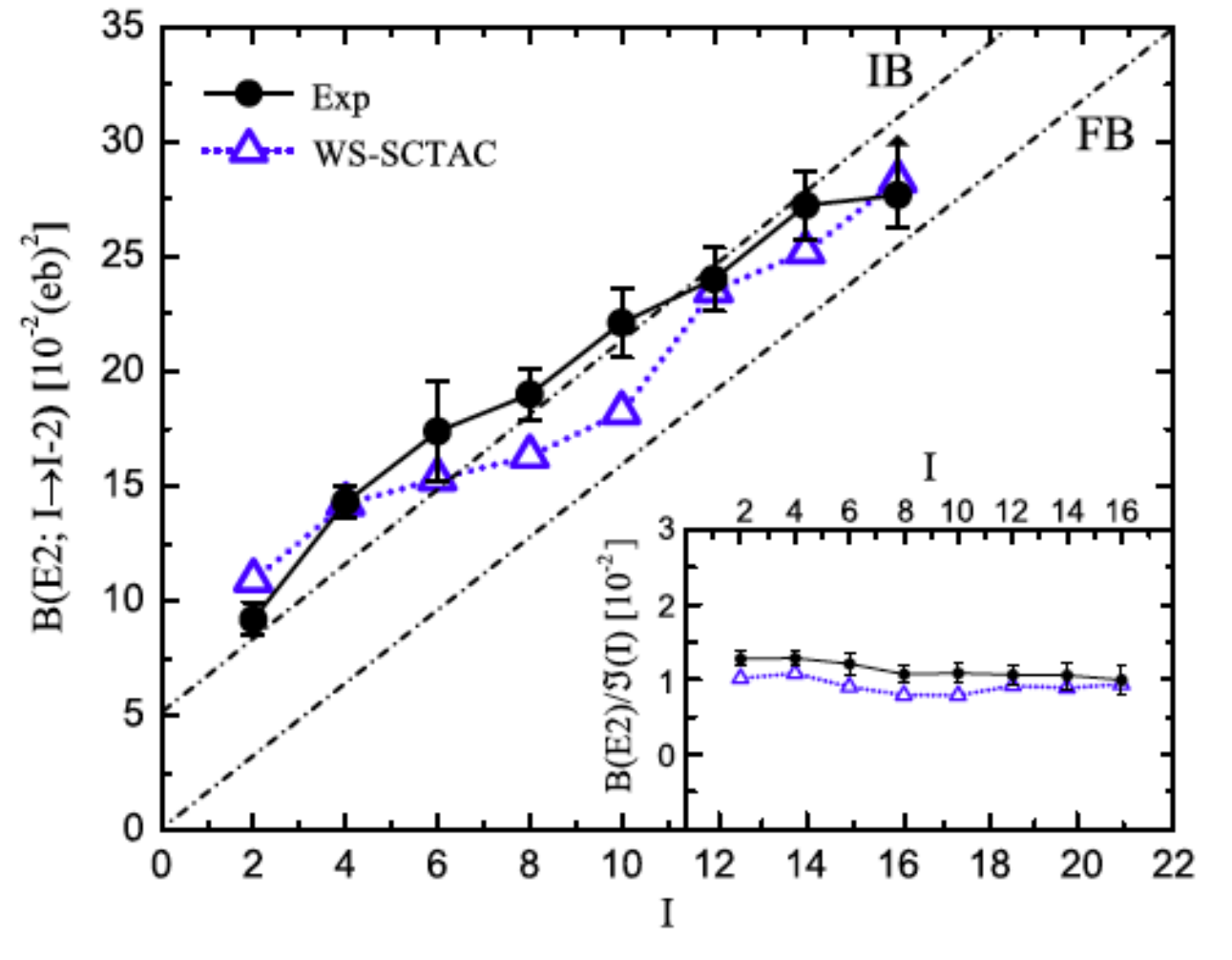,width=6.1cm}
\caption{\label{fig:102PdPRL} 
Left part: Experimental moments of inertia of the ground band 
of $^{102}$Pd, where \mbox{$\omega=(E(I)-E(I-2))/2$} and $J=I$.  Right part:  Experimental $B(E2, I\rightarrow I-2)$ values of $^{102}$Pd.
The blue lines WS-SCTAC show the calculations by means of the cranking model \cite{FGS11}. From Ref. \cite{102PdTidalPRL}.  } 
\end{center}
\end{figure}

Semiclassically,  $B(E2, I\rightarrow I-2) \propto Q_t^2$, because the transition quadrupole moment 
$Q_t =Q_2 \propto \beta_e \sin\gamma_e$ according to Eq. (\ref{eqn-bohr-q}). Therefore the ratio 
$B(E2)/{\cal J}$ does not depend on $J(=I)$. This correlation is well known as Grodzin's rule for the 2$^+_1$
states. It also applies with good accuracy for the $I>2$ yrast states  in nuclei with $44\leq Z\leq 48$ as far as the lifetimes 
of the states have been measured.  Fig. \ref{fig:102PdPRL}
shows $^{102}$Pd as an example.  Other regions have not been systematically investigated in this respect, though it holds for $^{154}$ Gd and $^{182}$Pt. 

Ayangeakaa {\it et al.}\cite{102PdTidalPRL} and Macchiavelli {\it et al.} \cite{MAFGC14} interpreted the classical tidal wave as a condensate of $d$ bosons. 
Accordingly, up to seven bosons are observed, which align their angular momenta. As discussed in Section \ref{sec:BNcut}, the structure of these $d$ bosons
must be more complicated than pairs of valence particles in a spherical potential coupled to angular momentum of two. If the bosons were free, the function 
${\cal J}(I)$ would be a straight line out of the coordinate origin (FB in Fig. \ref{fig:102PdPRL} and   V in Fig. \ref{fig:NdJIpots}). The displacement by $\Theta_0$  
was attributed  to an interaction between the bosons (IB FB in Fig. \ref{fig:102PdPRL}) that is quadratic in the boson number. From the systematics in the region
$44\leq Z\leq 48$  a correlation was 
identified: the smaller the boson energy the larger is the boson interaction.

\subsubsection{Semiclassical microscopic calculations.}\label{sec:TidalMicro}

Semiclassically, the yrast states of the quadrupole mode are represented by a tidal wave that  travels with constant angular 
velocity $\omega$ over the nuclear surface. This has allowed Frauendorf, Gu,and Sun \cite{FGS10,FGS11} to calculate its properties by means of the SC-TAC model (for details see \cite{TAC00}).
The potential $V(\beta,\gamma)$ in Eq. (\ref{Ebeta}) is determined by the Micro-Macro  method described in Section \ref{sec:mic-mac}.
It is based on the mean field Hamiltonian $h_{NBCS}$, which has the structure of $h_{PQQ}$ given by  Eq. (\ref{hPQQ}), combining the Nilsson deformed potential with the monopole pair field.  
 The rotational energy in Eq. (\ref{Ebeta}) is obtained by the cranking procedure 
\begin{eqnarray} 
\frac{J^2}{2{\cal J}(\beta,\gamma)}=E'(\omega,\beta,\gamma)+\omega J-E'(0,\beta,\gamma), 
~~\left<\omega,\beta,\gamma\vert j_x\vert\omega,\beta,\gamma\right>=J \label{erot}\\
\left[h_{NBCS}(\beta,\gamma)-\omega j_x\right]\vert\omega,\beta,\gamma\rangle=E'(\omega,\beta,\gamma)\vert\omega,\beta,\gamma\rangle \label{hcrank}.
\end{eqnarray}
For each deformation grid point $\beta,\gamma$, the frequency $\omega(J)$ is adjusted to obey the angular momentum constraint (\ref{erot}). The mean field state 
$\vert\omega,\beta,\gamma\rangle$ is found by the solving the the quasiparticle eigenvalue problem (\ref{hcrank}) exactly. This is a crucial difference to the ATDMF, which 
takes the cranking term $\omega j_x$ into account by perturbation theory, resulting in the IB expression for the moment of inertia.  The equilibrium values $\beta_e,\gamma_e$
 are found by minimizing $E(J,\beta,\gamma$).
The transition quadrupole moment is obtained as the expectation value $\left<J,\beta_e,\gamma_e\vert Q_2\vert J,\beta_e,\gamma_e\right>$

\begin{figure}[t]
\begin{center}
\psfig{file=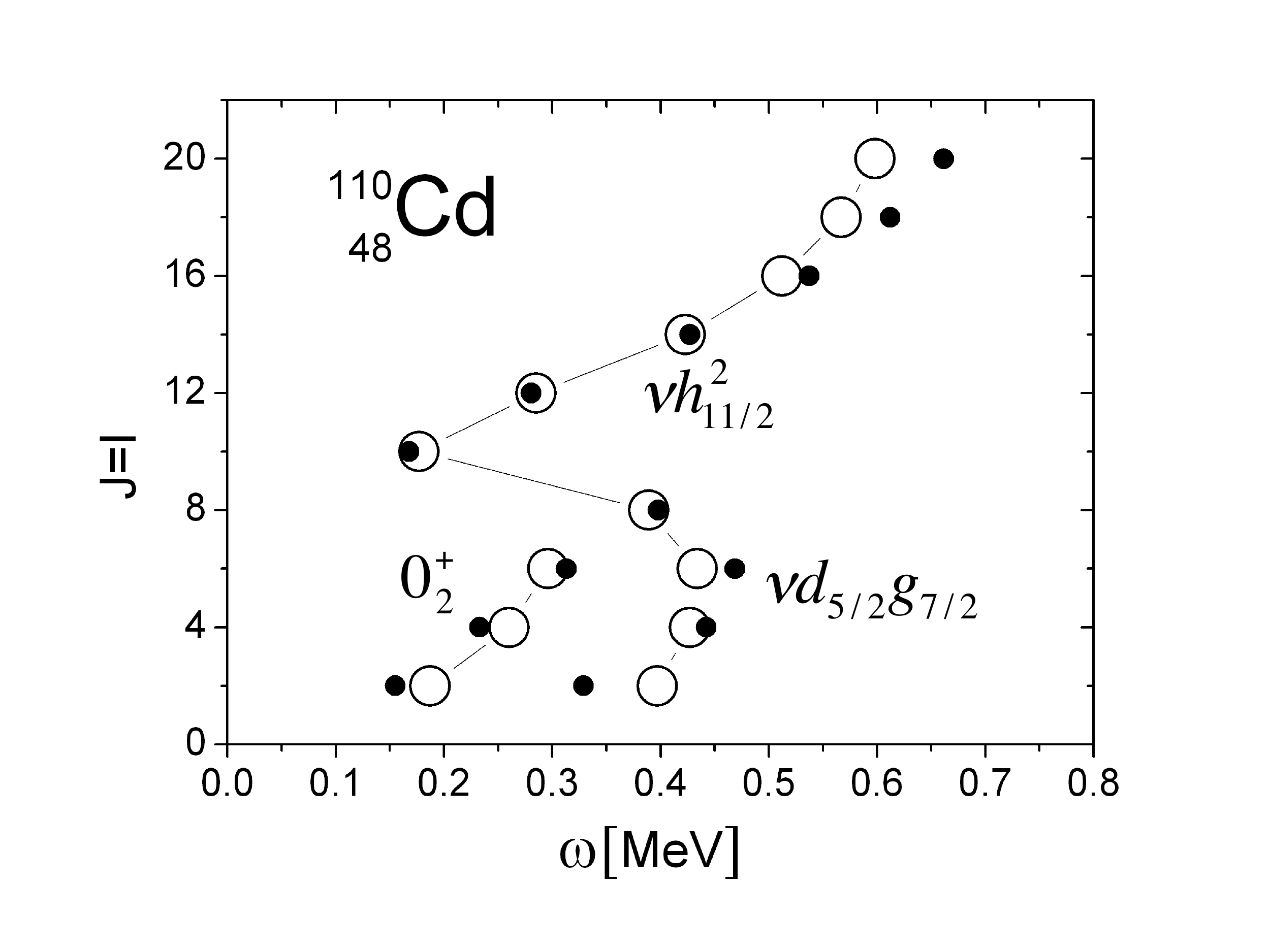,width=6.2cm}
\psfig{file=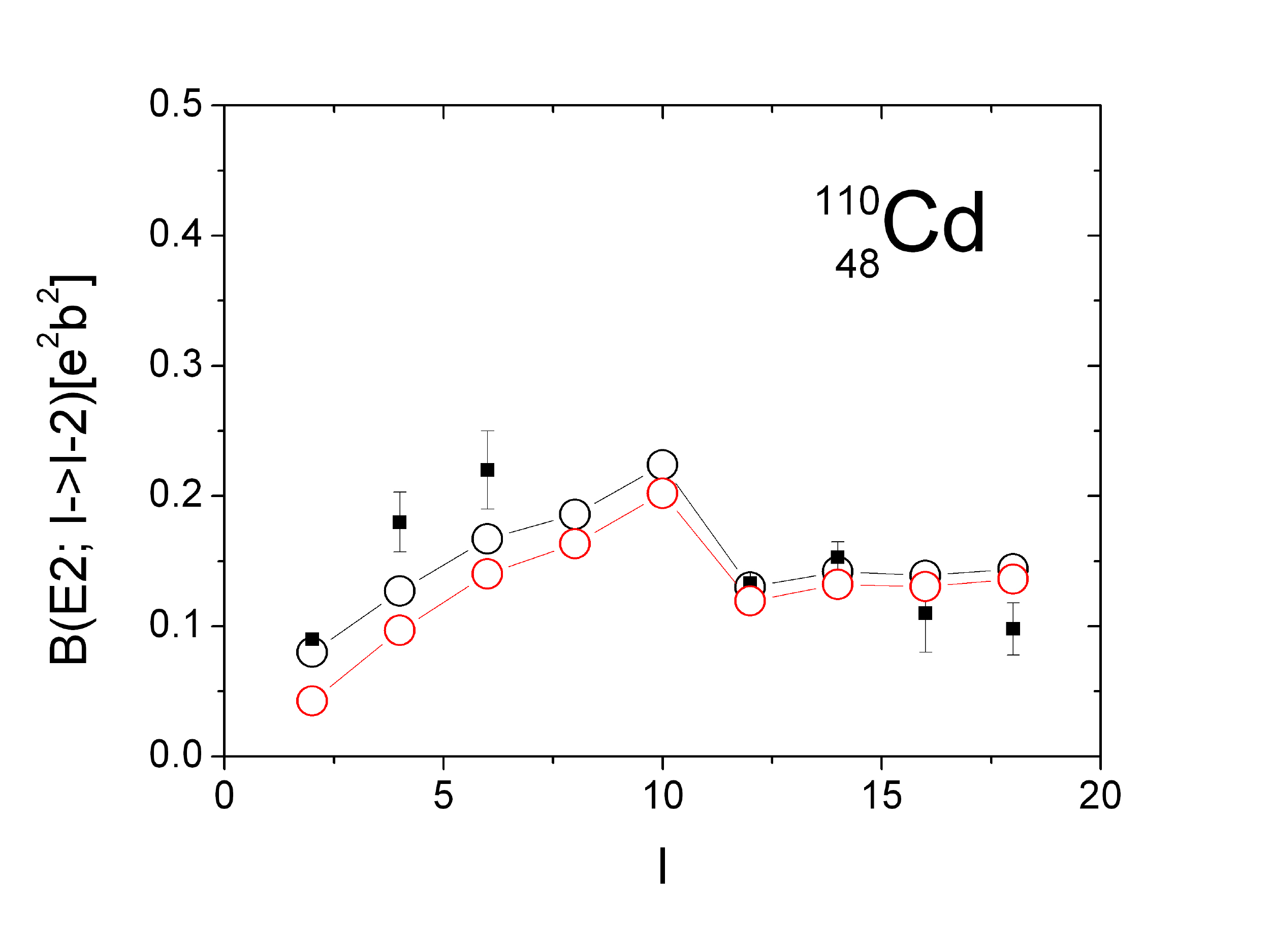,width=6.2cm}
\caption{\label{fig:110CdTAC} 
Left part: Experimental moments of inertia of the ground band and 0$^+_2$ band (black dots)
of $^{110}$Cd, where \mbox{$\omega=(E(I)-E(I-2))/2$} and $J=I$ compared with the  calculations by means of the Cranking model \cite{FGS11} (black circles).
 Right part:  Experimental $B(E2, I\rightarrow I-2)$ values of $^{110}$Cd compared with the  calculations (black circles without red circle with quantal correction, see Ref.
 \cite{FGS11}). From Ref. \cite{FGS11}.  } 
\end{center}
\end{figure}

Frauendorf, Gu,and Sun \cite{FGS10,FGS11} calculated the energies of the yrast states and the $B(E2)$ of the intra band transitions up to spin $I=16$ for the nuclides with $Z=44-48,~N=65-66$.  
The $g$-factors  for the same states were calculated in Ref. \cite{Chamoli11}, where the experimental deviations from  $Z/A$ could be reproduced. 
Figs. \ref{fig:102PdPRL} and \ref{fig:110CdTAC} exemplify the accuracy of the parameter-free calculations. In particular the change of the yrast states from the 
purely collective tidal wave (g band) to the  configuration  with two rotational aligned h$_{11/2}$quasiparticles  (s band) is reproduced in detail. 
In the case of $^{102}$Pd (Fig. \ref{fig:102PdPRL}) the collective g  band can be followed up to $I=14$, where it is at higher energy than the s   band. 
This is a consequence of almost no mixing of the two configurations.
In the case of $^{110}$Cd (Fig. \ref{fig:110CdTAC})  the collective g band is crossed by the s band earlier and the two bands interact stronger. The
two aligned h$_{11/2}$ quasiparticles  in the s band reduce the deformation but stabilizes it such that the sequence becomes more rotational. 
The rotational sequence based on the low-lying 0$^+_2$ state is also well reproduced. In the calculations it is a two-quasi proton excitation with larger deformation.
Its calculated energy is substantially larger than in experiment, which points to missing quadrupole correlations of vibrational type.
 More systematic calculations for other regions would be of interest. The method applies to odd-A and odd-odd  nuclei without any further sophistication. 

% \subsection{Triaxial Rotors}
 \subsection{Triaxial Projected Shell Model}
 Sheikh and  Hara \cite{SH99} introduced the Triaxial Projected Shell Model (TPSM) to describe the quadrupole mode in triaxial nuclei. The TPSM is based on the earlier axial version
 of the PSM, which has widely used for interpreting high spin experiments. Since this work has been reviewed in Ref. \cite{PSMrev},  only the more recent work on triaxial nuclei 
 will be covered here. The TPSM starts with a superposition of a set of quasiparticle configurations 
 $\vert\kappa\rangle$ in a triaxial Nilsson potential, which are projected by means of $P^I_{MK}$ on 
 good total angular momentum $I$ and its projection $K$ on one of the principal axes of the potential,

 \begin{figure}[t]
\begin{center}
\psfig{file=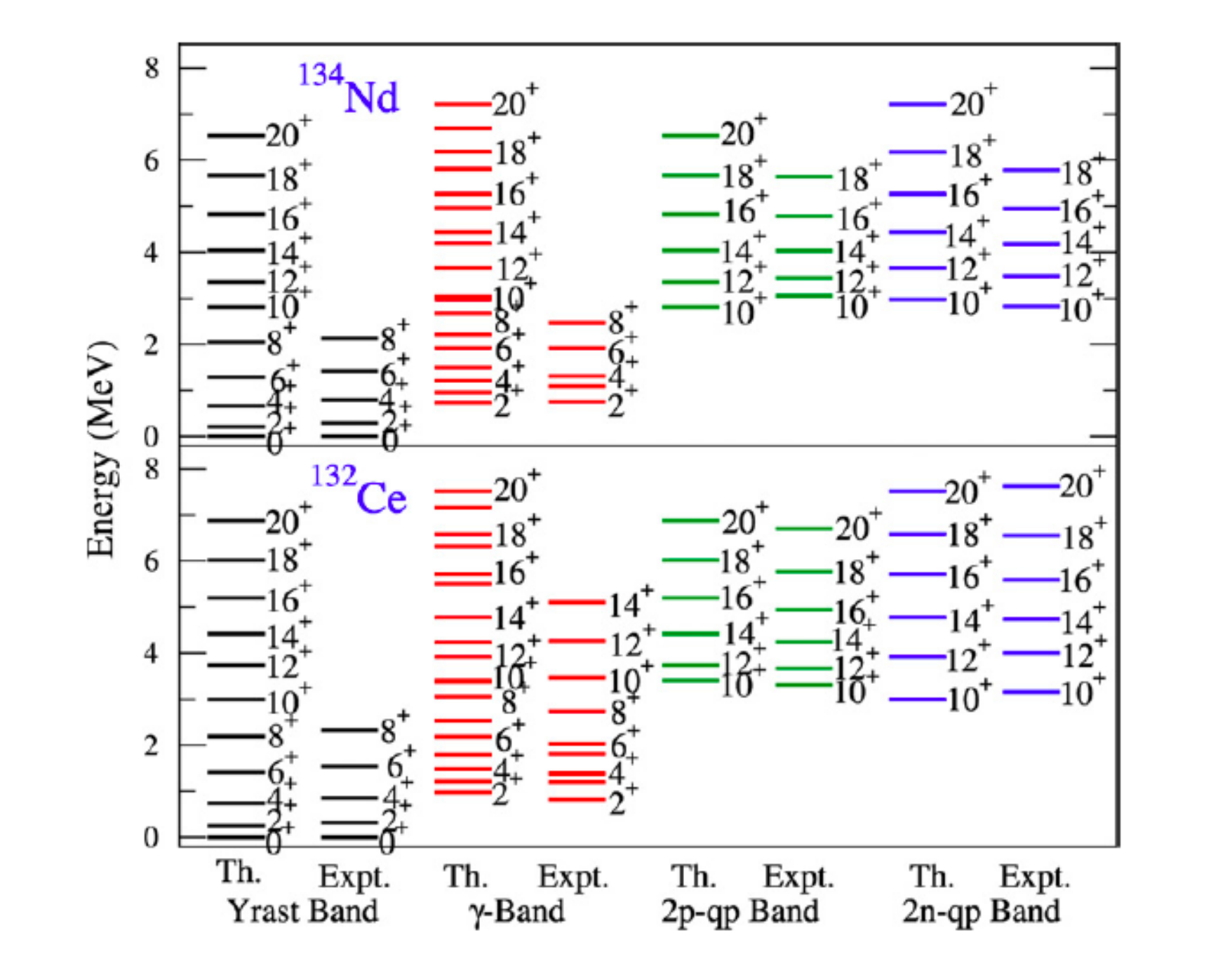,width=\linewidth}
\caption{\label{fig:134NdTPSM} The spectra of $^{134}$ Nd and $^{132}$Ce compared with TPSM calculations.
Reproduced with permission from Ref. \cite{Sheikh11Ce}.  } 
\end{center}
\end{figure}    
 
  \begin{eqnarray}
 \vert IM\sigma\rangle=\sum_{\kappa K}f^\sigma_{\kappa K}P^I_{MK}\vert \kappa\rangle,\\
 \sum_{\kappa' K'}{\cal H}_{\kappa K,\kappa' K'}f^\sigma_{\kappa' K'}=E_\sigma  \sum_{\kappa' K'}{\cal N}_{\kappa K,\kappa' K'}f^\sigma_{\kappa' K'},\\
 {\cal H}_{\kappa K,\kappa' K'}=\left< \kappa \vert H_{PQQ}P^I_{KK'}\vert\kappa' \right>,~~{\cal N}_{\kappa K,\kappa' K'}=\left< \kappa \vert P^I_{KK'}\vert\kappa'\right>.
 % g^\sigma_{\kappa K}= \sum_{\kappa' K'}{\cal N}_{\kappa K,\kappa' K'}^{1/2}f^\sigma_{\kappa' K'} 
 \end{eqnarray}
The amplitudes $f^\sigma_{\kappa  K}$ are found by diagonalizing the Pairing plus Quadrupole Hamiltonian $H_{PQQ}$ Eq. (\ref{HPQQ}) in the non-orthogonal basis.
They represent the collective quadruple mode of the triaxial rotor, which can be excited into multi-quasiparticle configurations.  
The axial deformation parameter $\beta$ is taken either from the measured $B(E2,~ 2^+\rightarrow0^+)$ values or calculated in the framework of some mean field approach.
It determines the coupling constant $\kappa$ of the QQ-interaction. The triaxiality parameter $\gamma$ is adjusted to reproduce the energy of the $2^+_2$ state (head of the quasi-$\gamma$ band). 
Values close to 30$^o$ are found for all cases. 

The coupling of the quadrupole mode to the quasiparticle excitations is a new quality. The strong coupling between the collective and quasiparticle degrees of freedom leads
 to new physics, as e.g. the appearance of chirality in $^{128}$Cs, which can be described in the framework of the TPSM \cite{Bhat12Cs}.  The capability of 
 TPSM to incorporate the coupling to the  quasiparticle degrees of freedom makes it an efficient tool to describe the structure of odd-A and odd-odd  nuclei
 on a microscopic basis. Examples are the studies of $^{103,105}$Rh in Ref. \cite{Bhat14Rh}  and $^{103}$Nb in Ref. \cite{Sheikh10Nb}
 
 Sheikh {\it et al.} studied in detail the properties of the $\gamma$ band in the  isotope chain $^{156-172}$Er in Refs. \cite{Sun00Er,Sheikh08Er,Sheikh11Er,Bout02Er}. 
  The calculations very well reproduce both energies and intraband $B(E2)$ values.  The work of Boutachkov {\it et al.} \cite{Bout02Er}
 demonstrates that TPSM also reproduces  the transition probabilities between the $\gamma$ and ground state band rather well. 
 The calculations include two-quasineutron and two-quasiproton excitations and the combinations thereof. This makes it possible to describe the 
 alignment of high-j quasiparticles on top of a $\gamma$ vibration, which is out of reach of other approaches.   In the TPSM framework,
the low-lying K = 3 bands observed in these nuclei are interpreted as built on triaxially deformed two-quasiparticle states \cite{Sheikh11Er}.  

 TPSM is predestinate for describing triaxial nuclei. Ref. \cite{ Bhat12Pt} investigated $^{180-190}$Pt and suggested that the low-lying 0$^+_2$ state 
 is a two-quasiproton excitation. This is in contrast to earlier interpretations which assign an axial shape to it that differs from the ground state shape.   
 The rigid-rotor like spectrum of $^{114}$Ru is very well accounted for by the calculations in Ref. \cite{Yeho11Ru}. The band structures of $^{128-134}$Ce 
 and $^{132-138}$Nd are accurately reproduced in the TPSM calculations of  Ref.\cite{Sheikh11Ce}. Fig. \ref{fig:134NdTPSM} shows how the 
 calculations describe both the collective ground and $\gamma$  bands and the high spin bands built on configurations with a rotationally aligned
 proton or neutron pair.  The study established a collective excitation of the $\gamma$-vibration type built on such two-quasineutron configuration, 
 which explains the negative $g$ -factor of the 10$^+$ state.
  Ref.\cite{Bhat14Ge} addressed 
  the even-odd staggering  of the $\gamma$ - band in $^{70-80}$Ge and $^{76-72}$Se.  
  As discussed in Sect. \ref{sec:triaxiality}, the 
  phase of the staggering parameter $S(I)$ indicates if the nucleus is $\gamma$-soft or rigid in the framework of the BH phenomenology. 
  The TPSM calculation  reproduces the surprising observation that
  $S(I)$ has the low-odd-$I$ character of $\gamma$-stiff nuclei for $^{76}$Ge while all neighbors have the low-odd-$I$ character of $\gamma$-soft nuclei.  
  
  The reason for the success of the TPSM in accounting for the details of the quasi-$\gamma$ bands remains to be understood. 
  The  quasi-$\gamma$  vibration can be
 understood as a wave  traveling the axial-symmetric surface of the nucleus, very much like the tidal wave the surface of a  spherical nucleus (cf. Sec. \ref{sec:tidal}).
 This corresponds to a static triaxial deformation in the co-rotating frame of reference. However, it seems counterintuitive that
 all considered cases, encompassing the relatively high-lying $\gamma$ bands 
 in the  well deformed Er isotopes and the low-lying quasi-$\gamma$ bands in $\gamma$-soft transitional nuclei with both phases of the staggering parameter $S(I)$,
 are well described by a deformed mean field with $\gamma\approx30^o$.  The properties of the  quasi-$\gamma$ bands do not correlate with 
 the mean field value of $\gamma$ in any obvious way, which is in stark contrast to the BH phenomenology.
    Moreover, a certain admixtures of two-quasiparticle configurations is needed to describe the experiment.
 
 The authors show the energies  $\left< \kappa \vert H_{PQQ}P^I_{KK}\vert\kappa \right>$ of the basis states before mixing, which they call "band diagrams".
 However the insight provided by these graphs is limited, because the properties of interest often emerge  as the consequence of the strong mixing a number of
 such basis states.  A more sophisticated analysis of the TPSM wave functions seems desirable to understand the physics behind the TPSM results.

\section{Assessment and challenges}\label{sec:assessment}

Basing the phenomenology  on the BH with irrotational flow type mass parameters and a forth oder potential (GCM) provides a classification 
scheme that qualitatively accounts for many features of the quadrupole mode in terms of three parameters. The model provides an illuminating interpretation in terms
of the nuclear shape for data in the whole range between the text book limits of harmonic vibration, rigid rotation (axial and triaxial), and  $\gamma$ instability.
The same holds for the IBM phenomenology, which has become popular because of the availability and simple use of the IBA codes. A similar handy  GCM 
code for public use would be very welcome. Both models are used to describe  the properties of the 
experimental  $2^+_1$, $4^+_1$, $6^+_1$, $2^+_2$, $3^+_1$, $4^+_2$, $0^+_2$, $4^+_3$ states, which they reproduce with comparable quality.
The often claimed foundation of the IBM on the microscopic Shell Model cannot be substantiated. The boson counting rule, which follows from this assertion,
has not been systematically tested, and it fails for $^{102}$Pd. The derivation of the IBM parameters using truncation schemes of the Shell Model configuration space have not
succeeded so far. Instead mapping of the potential energy surface calculated from the deformed mean field to the surface generated by the coherent state representation of IBM
 results in parameters that well account for the experiment.

The microscopic form of the BH is obtained by applying the adiabatic approximation to the TDMF theory. There are several versions that start from the various mean field
approaches used in practice at present, which are carefully tuned to experimental quantities other than the collective quadrupole excitations. 
The potential energy of the BH can be considered as reliable as the mean field approaches reproduce the properties of the single particle levels near the Fermi surface.
In this respect, the modern versions based on the Skyrme Energy Density Functional, the Relativistic Mean Field, and the Gogny Interaction  are not superior to
the older ones that start from a Modified Oscillator or Woods Saxon potential, which are directly adjusted to experimental properties of  the single particle levels.
Not surprisingly, the accuracy of the different approaches is comparable, because it cannot be better than the underpinning mean field theory.
All versions but one use the IB form for the mass parameters, which gives too small values for the mean field treatment of the pair correlations, resulting in a too
dilute excitation spectrum.  The discrepancy is removed by either simply scaling the spectrum or reduction of the pair correlations. 
The models perform well for the  yrast and yrare states with $I<6$ with considerable predictive power.  The performance is poorest for the excited 0$^+$ states, which
are expected to couple strongest to the quasiparticle states. Nevertheless, low-lying   0$^+$ states with shape coexistence, vibrational character or 
intermediate character   are quite well accounted for. The pronounced deformation dependence of the mass parameters is essential for establishing these structures.
Calculating by means of the microscopic BH the spectroscopic properties of the quadrupole excitations across the mass table  is feasible with modern computers. 
Such a project has been only carried out for the version based on the Gogny interaction, the results of which is accessible online. To create a similar resource for the other versions 
of the microscopic BH would be very useful, both for interpreting experiments and for a more systematic assessment of the model performance.

Taking into account the time-odd
parts of the quadruple pair field by means of the LQRPA method increases the mass parameters from the IB to the TV values, which give the correct energy scale. 
The LQRPA is computationally demanding, and it has only been applied for the simple PQQ interaction for this reason. Developing tractable methods to calculate the TV 
mass parameters for the other types of interactions or EDF's is a challenge. 

The Generator Coordinate Method (without GOA) applied to all five quadrupole degrees of freedom is computationally very extensive. For this reason it has been restricted to
very light nuclei. The method is particularly well suited for parallel computing, which may lead to progress with the further development of super computers.  
For the heaviest nucleus studied, $^{76}$Kr, it gives similar result as the microscopic BH derived from the same RMF EDF.   
So far the collective states have been generated from the ground state configurations on the quadrupole coordinate grid, which implicitly invokes the adiabatic approximation.
The coupling to quasiparticle excitation can be incorporated in a straight forward way, such going beyond the adiabatic regime. To make this tractable is certainly a challenge.

A wide stride into this direction is the Triaxial Projected Shell Model, which assumes a fixed deformation, 
projects on good angular momentum, and takes coupling to up to four-quasiparticle configurations
into account. Using the simple PQQ interaction makes the calculations practical. The method includes an element of phenomenology, 
because the shape of the mean field is partially adjusted to reproduce the experiment.    
 The coupling between the collective quadrupole  and the quasiparticle  excitations  is remarkable well reproduced
for the states in the yrast region. This is
also true for odd-A and odd-odd nuclei, to which the method applies without any complications.  Often the results are obtained  as a heavily mixed 
 number of components in a non-orthogonal basis.
It is a challenge  to find interpretations of such results in simple terms in oder to elucidate the physics behind them. Treating the deformation parameters  as generator coordinates appears an 
attractive future development. The simplicity of the PQQ interaction will keep the computational requirements as low as possible, which facilitates the exploration of the new territory.    
The not-yet-understood observation that the model  describes the various modes of the quasi-$\gamma$ band ($\gamma$-soft vs. $\gamma$-rigid) assuming
one static mean field  triaxiality of $\gamma\approx 30^o$ may  considerably  simplify the calculations.       
 
The Tidal Wave concept  provides a simple semiclassical description for the yrast states.     The interplay between the collective quadrupole mode and the quasiparticles is 
described by quasiparticle configurations in a uniformly rotating potential. The method works for the whole range of vibrational to rotational nuclei, for which it is well tested and 
used as the standard classification of the rotational bands.  The method very well reproduces the energies of the yrast levels and the $B(E2)$ values of the 
connecting transition for the Ru, Pd, and Cd isotope chains. It can be directly applied to odd-A and odd-odd nuclei.  The semiclassical approximation is not quite accurate  for the low- spin 
states $I<6$. Projecting on good angular momentum should remove this deficiency. 
 
Exploring the region of strong coupling  between the collective and quasiparticle degrees of freedom will require a paradigm change. Some large-dimension diagonalization has to be used
to account for the rapid increase of the level density. The nucleus is a partially chaotic system. Only close to the yrast line one can expect that the theory reproduces the experiment 
state by state, and the traditional comparison of individual energies and transition probabilities is appropriate.  Higher up one has to take into account a certain degree of randomness and 
study the properties of groups of levels. It is a challenge to formulate new concepts that identify the correlations due to the fragmented quadrupole mode as long as it survives.
This concerns also standard Shell Model calculations like the example in Fig. \ref{fig:decoherence}. It is not clear from simple plot of the $B(E2)$ values of the connecting transitions
whether the low-spin members of the three- and higher phonon multiplets survive in a fragmented form or if the quadrupole collectivity is quenched. The availability of the full information about 
the wave function allows theory to address this question in novel ways.

\section*{Acknowledgements}
The work was supported by the U.S. DOE under Contract Nos. DEFG02- 95ER-40934.

\end{document}